\documentclass[a4paper,11pt]{article}
\pdfoutput=1 

\usepackage{jheppub} 

\usepackage[T1]{fontenc} 

\def\gsim{\raise0.3ex\hbox{$\;>$\kern-0.75em\raise-1.1ex\hbox{$\sim\;$}}}
\def\lsim{\raise0.3ex\hbox{$\;<$\kern-0.75em\raise-1.1ex\hbox{$\sim\;$}}}

\def\vev#1{\left\langle #1\right\rangle}

 \newcommand{\cy}[1]{{\color{cyan}  #1}}
\newcommand {\ignore}[1]{}

\newcommand{\sm}{{Standard Model }}

\def\SM{$\mathrm{SU(3)_c \otimes SU(2)_L \otimes U(1)_Y}$ }

\def\321{$\mathrm{SU(3) \otimes SU(2) \otimes U(1)}$ }

\def\black{\color{black}{}}

\def\cy#1{\textcolor{cyan}{#1}}

\newcommand{\AddrAHEP}{
  AHEP Group, Institut de F\'{i}sica Corpuscular --
  CSIC/Universitat de Val\`{e}ncia, Parc Cient\'ific de Paterna.\\
 C/ Catedr\'atico Jos\'e Beltr\'an, 2 E-46980 Paterna (Valencia) - SPAIN}
 
 \newcommand{\AddrIISERB}{Department of Physics,
 Indian Institute of Science Education and Research - Bhopal \\
 Bhopal Bypass Road, Bhauri, Bhopal, India}

\title{Electroweak symmetry breaking in the inverse seesaw mechanism}
\author[a]{Sanjoy Mandal,}
\author[b]{Rahul Srivastava,}
\author[a]{Jos\'{e} W. F. Valle}
\affiliation[a]{\AddrAHEP}
\affiliation[b]{\AddrIISERB}

\emailAdd{smandal@ific.uv.es}
\emailAdd{rahul@iiserb.ac.in}
\emailAdd{valle@ific.uv.es}

\abstract{We investigate the stability of Higgs potential in inverse seesaw models.
    We derive the full two-loop RGEs of the relevant parameters, such as the quartic Higgs self-coupling, taking thresholds into account.
    We find that for relatively large Yukawa couplings the Higgs quartic self-coupling goes negative well below the \sm instability scale $\sim 10^{10}$ GeV.
    We show, however, that the ``dynamical'' inverse seesaw with spontaneous lepton number violation can lead to a completely
    consistent and stable Higgs vacuum up to the Planck scale.}
\begin{document} 
\maketitle
\flushbottom


\section{Introduction}
\label{sec:introduction}

\begin{figure}[h]
\centering
\includegraphics[width=0.75\textwidth]{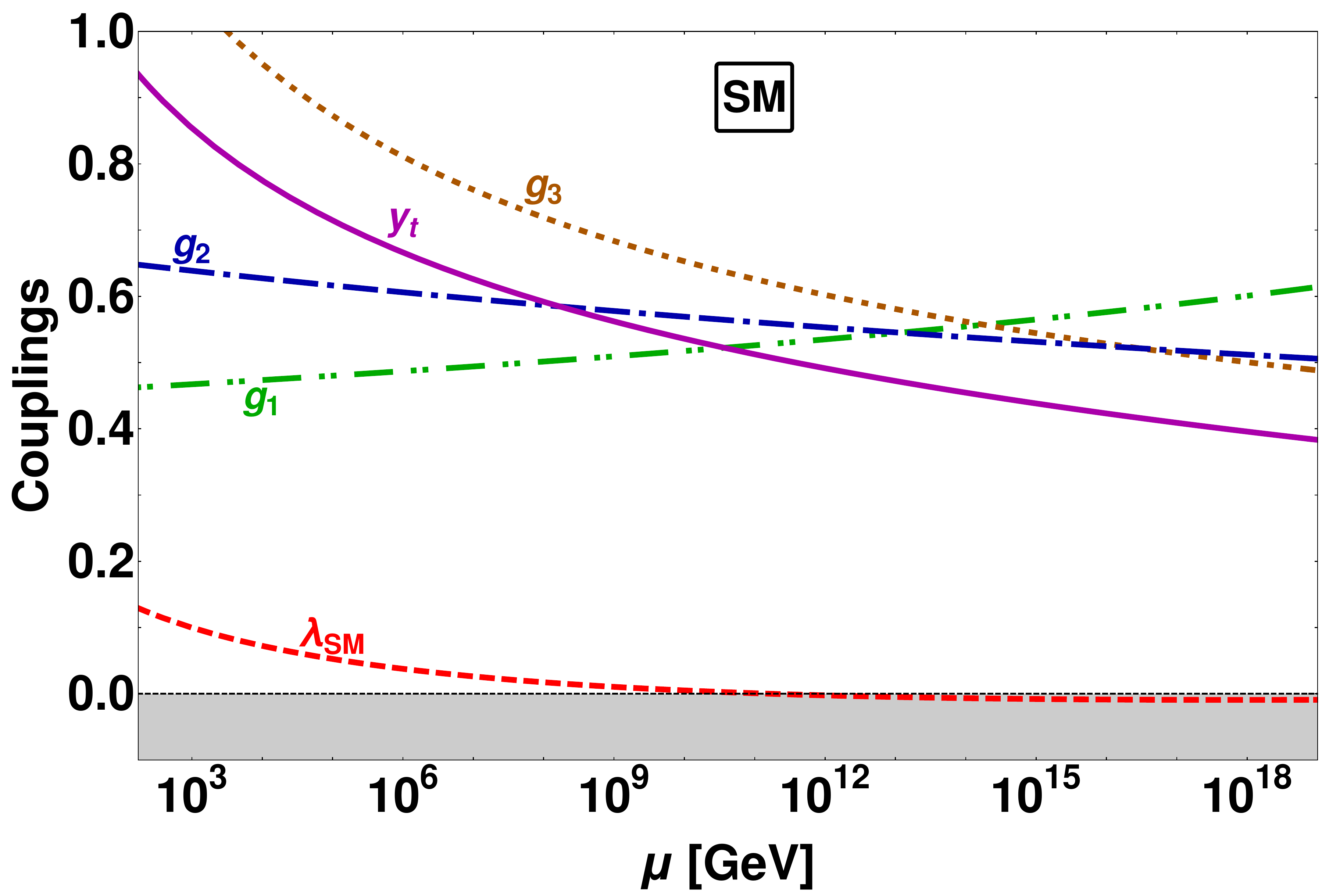}
\caption{\footnotesize{The renormalization group evolution of the \sm gauge couplings $g_{1}$, $g_{2}$, $g_{3}$, the top quark Yukawa coupling $y_{t}$ and the quartic Higgs boson self-coupling
    $\lambda_{\text{SM}}$.
      Here we adopt the $\overline{\text{MS}}$ scheme, taking the parameter values at low scale as input, see \cite{Mandal:2019ndp} for details.} }
\label{fig:RG-SM-Running}
\end{figure}

   The historical discovery of the Higgs boson~\cite{Aad:2012tfa,Chatrchyan:2012ufa} and the subsequent precise measurements of its properties~\cite{Tanabashi:2018oca} can be used to shed light on the
   electroweak symmetry breaking mechanism.
   In particular, we can now not only determine the value of the quartic coupling of the \sm scalar potential at the electroweak scale, but also use it to shed light on possible new physics all the way up to Planck scale.
   Given the present measured top quark and Higgs boson masses, one can calculate the corresponding Yukawa $y_t$ and Higgs quartic $\lambda_{\text{SM}}$ couplings within the Standard Model.
     These, along with the \SM gauge couplings $g_1, g_2, g_3$ respectively, are the most important input parameters characterizing the \sm renormalization group equations (RGEs).
     Given the values of these input parameters\footnote{The numbers given in Table~\ref{tab:inputs} are the central values.
        We use them as the input parameters for our RGEs.
         The importance of errors has been studied in Ref.~\cite{Buttazzo:2013uya}, to which we refer the reader for more details.}, as shown in Table~\ref{tab:inputs}, the Higgs quartic coupling tends to run negative between the electroweak and Planck scales, as seen in Fig.~\ref{fig:RG-SM-Running}. 
\begin{table*}[ht]
	\centering
	\begin{tabular}{|c|c|c|c|c|c|}
		\hline
		                 &   $g_{1}$  &  $g_{2}$  &  $g_{3}$  & $y_{t}$  &  $\lambda_{\text{SM}}$  \\
		\hline
		$\mu (m_t)$      &   0.462607                   &  0.647737 &  1.16541  & 0.93519 &  0.126115   \\
		\hline
	\end{tabular}
        \caption{\footnotesize{
                $\overline{\text{MS}}$ values of the input parameters at the top quark mass scale, $\mu( m_{t}) = 173\pm 0.4$ GeV~\cite{Tanabashi:2018oca}.    }
        }
	\label{tab:inputs}
\end{table*}

  One sees that the \sm Higgs quartic coupling $\lambda_{\text{SM}}$ becomes negative at an energy scale $\sim 10^{10}$ GeV. 
This would imply that the \sm Higgs potential is unbounded from below. 
   Hence, the \sm vacuum is not absolutely stable~\cite{Degrassi:2012ry,Alekhin:2012py,Buttazzo:2013uya}.
   Instead, these next-to-next-to-leading order analyses of the \sm Higgs potential suggest that the vacuum is actually metastable.\\[-.2cm]

 Moreover, despite its many successes, the \sm cannot be the final theory of nature.
One of its main shortcomings is its inability to account for neutrino mass generation, needed to describe neutrino oscillations~\cite{deSalas:2020pgw}.
   The Higgs vacuum stability problem in neutrino mass models can become worse than in the \sm~\cite{Khan:2012zw,Rodejohann:2012px,Bonilla:2015kna,Rose:2015fua,Lindner:2015qva,Ng:2015eia,Bambhaniya:2016rbb,Garg:2017iva}.
   Here we follow Ref.~\cite{Mandal:2019ndp} and confine ourselves to the Standard-Model-based seesaw mechanism using the simplest \SM gauge group.\\[-.2cm]

 The latter can be realized in ``high-scale'' schemes with explicit~\cite{Schechter:1980gr} or spontaneous violation of lepton number~\cite{Chikashige:1980ui,Schechter:1981cv}.
   These typicaly involve messenger masses much larger than the electroweak scale.
  Alternatively, neutrino mass may result from ``low-scale'' physics ~\cite{Boucenna:2014zba}.
For example, the type-I seesaw mechanism can be mediated by ``low-scale'' messengers. This happens in the inverse seesaw mechanism.
 Lepton number is broken by introducing extra \SM singlet fermions with small Majorana mass terms, in addition to the conventional ``right-handed'' neutrinos.
 Again, one can have either explicit~\cite{Mohapatra:1986bd} or spontaneous lepton number violation~\cite{GonzalezGarcia:1988rw}.  \\[-.2cm]
%
\begin{figure}[h]
\centering
\includegraphics[width=0.6\textwidth]{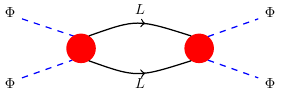}
\caption{\footnotesize{Destabilizing effect of Weinberg's effective operator on the Higgs quartic interaction. }}
\label{dimension-five}
\end{figure}

Any theory with massive neutrinos has an intrinsic effect, illustrated in Fig.~\ref{dimension-five}, that may potentially destabilize the electroweak vacuum~\footnote{In the presence of very specific symmetries this model-independent argument might be circumvented.}.
  This vacuum stability problem becomes severe in low-scale-seesaw schemes ~\cite{Mandal:2019ndp}.
  Indeed, if the heavy mediator neutrino lies in the TeV scale, its Yukawa coupling will run for much longer than in the high-scale type-I seesaw.
As a consequence, the quartic coupling $\lambda$ tends to become negative sooner, much before the \sm instability sets in.

  Here we examine the consistency of the electroweak symmetry breaking vacuum within the inverse seesaw mechanism.
Apart from the destabilizing effect illustrated in Fig.~\ref{dimension-five} there will in general be other, model-dependent, and possibly leading contributions that can reverse this trend.
  We note that the spontaneous violation of lepton number,
  implying the existence of a physical Nambu-Goldstone boson, dubbed majoron~\cite{Chikashige:1980ui,Schechter:1981cv}, can substantially improve the electroweak vacuum stability properties.
  Indeed, the extended scalar sector of low-scale-majoron-seesaw schemes plays a key role in improving their vacuum stability.
%
%
 This sharpens the results presented in Ref.~\cite{Bonilla:2015kna}.
Indeed, we find that renormalization group~(RG) evolution can cure the vacuum stability problem in inverse seesaw models
also in the presence of threshold effects. These can be associated both with the scalar as well as the fermion sector of the theory~\footnote{Notice that, while Ref.~\cite{Mandal:2019ndp} included threshold effects, in the high-scale seesaw framework such effects appear only at high energies, and do not affect low-scale physics.}. 

  The paper is organized as follows. 
  In Section~\ref{sec:inverse-sees-mech}, we describe neutrino mass generation in the inverse-seesaw model.
  In Section~\ref{sec:higgs-vacu-stab} we show that the vacuum stability problem becomes worse within the simplest inverse-seesaw extensions with explicitly broken lepton number.
  In Section~\ref{sec:major-compl-inverse}, we then focus on the majoron completion of the inverse seesaw.
  We then show in Section~\ref{sec:vacu-stab-inverse} how the majoron helps stabilize the Higgs vacuum, all the way up to Planck scale.
In Section~\ref{sec:comp-stand-miss}, we compare the vacuum stability properties of the various missing-partner-inverse-seesaw variants with those of the sequential case.
%
%
In Sec.~\ref{sec:invisible} we briefly illustrate the interplay between vacuum stability and the restrictions on the Higgs boson invisible decays~\cite{Joshipura:1992hp} that follow from current
  LHC experiments.
  Finally, we conclude and summarize our main results in Section~\ref{sec:Conclusions}.


\section{The Inverse Seesaw mechanism}
\label{sec:inverse-sees-mech}


 The issue of vacuum stability must be studied on a model-by-model basis.
  In this work we examine it in the context of inverse-seesaw extensions of the Standard Model.
  The inverse seesaw mechanism is realized by adding two sets of electroweak singlet ``left-handed'' fermions $\nu^c_{i}$ and $S_{i}$~\cite{Mohapatra:1986bd,GonzalezGarcia:1988rw}.
   The relevant part of the Lagrangian is given by 
\begin{align} 
 -\mathcal{L}=\sum_{ij} Y_{\nu}^{ij} L_{i}\tilde{\Phi} \nu^c_{j} + M^{ij} \nu^c_{i} S_{j}  + \frac{1}{2}\mu^{ij}_S S_{i} S_{j} + \text{H.c.} 
 \label{lag-inv-seesaw}
\end{align}
%
  where $L_{i} = \begin{pmatrix} \nu & \ell \end{pmatrix}^{T}$;$i = 1,2,3$ are the lepton doublets, $\Phi$ is the \sm Higgs doublet, $M$ is the Dirac mass term.
  The two sets of fields $\nu^c$ and $S$ transform under the lepton number symmetry $U(1)_L$ as $\nu^c \sim -1$ and $S \sim +1$, respectively.
  The $M$ and $\mu_S$ terms are both gauge invariant mass matrices, but only $M$ is invariant under lepton number symmetry, since $\mu_S$ violates lepton number by two units.
Light neutrino masses are generated through the tiny lepton number violation. Indeed, after electroweak symmetry breaking, the effective light neutrino mass matrix has the following form
\begin{align}
\mathcal{M}_{\nu}=
 \begin{pmatrix}
  0 & m_{D}  & 0 \\
  m_{D}^{T} & 0 & M \\
  0 &  M^T  &  \mu_{S}  \\
 \end{pmatrix},
 \label{Mass Matrix}
\end{align}
%
with $m_D = \frac{v}{\sqrt{2}} Y_\nu$. Neutrino masses arise by block-diagonalizing Eq.~\ref{Mass Matrix} as,
%
\begin{align}
 \mathcal{U}^T.\mathcal{M}_\nu.\mathcal{U}= \mathcal{M}_D
\end{align}
%
through the unitary transformation matrix $\mathcal{U}$, where $\mathcal{M}_D$ has a block-diagonal form. 
Since the lepton number is retored as $\mu_S \to 0$, the symmetry breaking entries of $\mu_S$ can be made naturally small in the sense of t'Hooft.
Apart from symmetry protection, the smallness of $\mu_S$ may also result from having a radiative origin associated to new physics such as supersymmetry,
left-right symmetry or dark matter physics~\cite{Bazzocchi:2009kc,CarcamoHernandez:2018hst,Rojas:2019llr}.
In contrast, being gauge and lepton-number invariant, the elements of $M$ are expected to be naturally large.
Thus we obtain the hierarchy $M\gg m_D\gg \mu_S$.
Under this hierarchy assumption we perform the standard seesaw diagonalization procedure~\cite{Schechter:1981cv}, to obtain the effective light neutrino mass matrix $m_\nu$ as
%
\begin{align}
 m_{\nu} \approx m_{D} M^{-1} \mu_S (M^{T})^{-1} m_{D}^{T} = \frac{v^{2}}{2} Y_{\nu} M^{-1} \mu_S (M^{T})^{-1} Y_{\nu}^{T}
\end{align}

Furthermore, in contrast to conventional type-I seesaw, the scale of lepton number violating parameter $\mu_S$ is much smaller than the characteristic mediators scale $M$.
As a result, the heavy singlet neutrinos become quasi-Dirac-type fermions~\footnote{The concept of quasi-Dirac fermions was first suggested for the light neutrinos in~\cite{Valle:1982yw}.
It constitutes a common feature of all low-scale seesaw models.}.
  Note that, the small lepton number violating Majorana mass parameters in $\mu_S$ control the smallness of light neutrino masses.
  As $\mu_S \to 0$, the global lepton number symmetry is restored, and as a result, all the three light neutrinos are strictly massless.
Small neutrino masses are ``symmetry-protected'' by the tiny value of $\mu_S \neq 0$.
The smallness of $\mu_S$ allows the Yukawa couplings $Y_{\nu}$ to be sizeable, even when the messenger mass scale $M$ lies in the TeV scale, without conflicting with the observed smallness of neutrino masses.

In contrast to the high-scale type-I seesaw, in inverse-seesaw schemes one can have a very rich phenomenology that makes them potentially testable in current or upcoming experiments.
  For example, the mediators would be accessible to high-energy collider experiments~\cite{Dittmar:1989yg,Abreu:1996pa,Acciarri:1999qj,Cai:2017mow},
  with stringent bounds, e.g. from the Delphi and L3 collaborations~\cite{Abreu:1996pa,Acciarri:1999qj}.
Moreover, they would induce lepton flavour and leptonic CP violating processes with potentially large rates,
unsuppressed by the small neutrino masses~\cite{Bernabeu:1987gr,Branco:1989bn,Rius:1989gk,Deppisch:2004fa,Deppisch:2005zm}. 
Finally, since the mediators would not take part in low-energy weak processes, the light-neutrino mixing matrix describing oscillations would be effectively
non-unitary~\cite{Valle:1987gv,Nunokawa:1996tg,Antusch:2006vwa,Miranda:2016ptb,Escrihuela:2015wra}.
%
%
In short, in contrast to the conventional high-scale seesaw, the inverse seesaw mechanism could harbor a rich plethora of accessible new physics processes, that could be just around the corner.

As $\nu^c$ and $S$'s are \sm gauge singlets, carrying no anomalies, there is no theoretical limit on their multiplicity.
  Many possibilities can arise depending on the number of $\nu^c$ and $S$ in a given model.
In the sequential inverse seesaw model the number of $\nu^c$ matches that of $S$, and there are three ``heavy'' quasi-Dirac leptons in addition to the three light neutrinos.
  For the case of different number of $\nu^c$ and $S$, in addition to the light and heavy neutrinos, the spectrum will also contain intermediate states with mass proportional to $\mu_S$. 
  These could be warm dark matter candidate if their mass lies in KeV scale~\cite{Abada:2014zra}. \\[-.2cm]

  For the sake of simplicity, here we consider only the case where $\nu^c$ and $S$ come with the same multiplicity.
  Moroever, since adding more fermion species will only worsen the Higgs vacuum stability problem, in section~\ref{sec:higgs-vacu-stab} we opt for the minimal (3,1,1) case, namely a single pair of lepton mediators.
  In such minimal ``missing-partner'' seesaw~\cite{Schechter:1980gr} two of the light neutrinos will be left massless.
  In Section~\ref{sec:comp-stand-miss} we examine the quantitative differences between the different multiplicity choices concerning the issue of vacuum stability. 
  Moroever, we briefly discuss the phenomenological viability of the various options.

\section{Higgs vacuum stability in inverse seesaw}
\label{sec:higgs-vacu-stab}

In the above preliminary considerations we have briefly summarized the main features of the inverse seesaw model.
We now examine the effect of the new fermions $\nu^c$ and $S$ upon the stability of the electroweak Higgs vacuum.
We take into account the effect of the thresholds associated with the extra fermions $\nu^c$ and $S$, as well as the scalars (in Section~\ref{sec:major-compl-inverse} and~\ref{sec:vacu-stab-inverse}) responsible for the spontaneous breaking of lepton number.

\subsection{Effective Theory}
\label{sec:effective-theory}

To begin with, in the effective theory where the heavy singlet fermions $\nu^c$ and $S$ are integrated out 
we 
have a natural threshold scale $\Lambda \approx M$ given by their mass, see Eq.~(\ref{lag-inv-seesaw}). 
%
  %
  As as a result, below this scale the theory is the \sm plus an effective dimension five Weinberg operator~\cite{Weinberg:1979sa}, given by 
%
\begin{align}
-\mathcal{L}_\nu^{d=5}=\frac{\kappa}{2}\,  L \, L \, \Phi \, \Phi \, + \,  \text{H.c.}
\label{eff-op}
\end{align}
%
where $\kappa = (Y_\nu M^{-1}\mu_S (M^{T})^{-1} Y_\nu^T)$ is the $3\times 3$ effective coupling matrix.
Unless they are needed, in what follows we will suppress the generation indices.
Note that $\kappa$ has negative mass dimension. The above Lagrangian leads to a left-handed neutrino Majorana mass matrix as
\begin{align}
m_\nu\equiv \kappa \frac{v^2}{2}
\end{align} 
As a result, below the scale $\Lambda$, only the \sm couplings and $\kappa$ will run.
  Neglecting lepton and light quark Yukawa couplings, the one-loop RGEs~\cite{Chankowski:2000fp,Antusch:2002rr,Bergstrom:2010id} are given by~\cite{Mandal:2019ndp}
\begin{align}
16\pi^2\beta_{\kappa}=6y_t^2\kappa-3g_2^2\kappa+\lambda_\kappa \kappa
\end{align}
Due to the large top Yukawa coupling, $\kappa$ slowly increases with the threshold scale $\Lambda$.
We denote the Higgs quartic coupling in this case as $\lambda_\kappa$ to distinguish it from the pure Standard Model case.
The above Weinberg operator also gives a correction to the Higgs quartic coupling $\lambda_\kappa$ below the scale $\Lambda$.
The contribution of the coupling $\kappa$ to the running of $\lambda_\kappa$ is of order $v^2\kappa^2$ and thus negligible, as shown in~\cite{Ng:2015eia,Bergstrom:2010id,Mandal:2019ndp}.
Hence, below the scale $\Lambda$, the evolution of $\lambda_\kappa$ will be almost the same as in the Standard Model.
\subsection{Full Theory}
\label{sec:full-theory}
We now turn to the region above the threshold scale  $\Lambda$. In this regime we have the full Ultra-Violet (UV) complete theory.
  Hence one must take into account the RGEs of all the new couplings present in the model, as they will affect the evolution of the Higgs quartic coupling.
  In particular, we will see that the stability of the electroweak vacuum limits how large the Yukawa coupling $Y_\nu$ can be.
  The Higgs quartic self-coupling in full UV-complete theory will be denoted by $\lambda$, to distinguish it from the \sm coupling $\lambda_{\text{SM}}$ and from the effective theory quartic
  coupling $\lambda_\kappa$ discussed above.  \\[-.2cm]

  For simplicity we will first study the case of just one species of $\nu^c$ and $S$, which we call the $(3,1,1)$ inverse seesaw.
  As mentioned, this of course is not -- by itself -- realistic, as in this case only one of the light neutrinos obtains mass.
  However, the missing mass parameter may arise from a different mechanism~\cite{Rojas:2019llr} associated, say, with dark matter.
  Moreover, the $(3,1,1)$ case provides the simplest reference scheme, that brings out all the relevant features.
  In Section~\ref{sec:comp-stand-miss} we will compare with the $(3,2,2)$ and the $(3,3,3)$ -- the sequential inverse seesaw mechanim -- with two and three species of $\nu^c$ and $S$, respectively.  
\begin{figure}[h]
\centering
\includegraphics[width=0.3\textwidth]{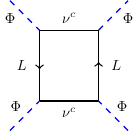}
\caption{\footnotesize{The destabilizing effect of right-handed neutrinos on the evolution of the Higgs quartic coupling.} }
\label{Yukawa negative effect}
\end{figure}

The running of $Y_\nu$ above the threshold scale is governed by the RGEs given in Appendix.~\ref{sec:rges:-inverse-seesaw}.
  Apart from the RG evolution, one must also take into account the threshold corrections, associated with integrating the heavy fermions in the effective theory.
The tree-level Higgs potential is given by
\begin{align}
V = -\mu_\Phi^2 (\Phi^\dagger \Phi) + \lambda (\Phi^\dagger \Phi)^2
\end{align} 
  This will get corrections from higher loop diagrams of Standard Model particles as well as from the extra fermions present in the inverse seesaw model.
  It introduces a threshold correction to the Higgs quartic coupling $\lambda$ at $\Lambda = M$.
  Here we follow Ref.~\cite{Mandal:2019ndp} in estimating this threshold correction as $\Delta\lambda_{\text{TH}}=-\frac{5}{32\pi^2}|Y_\nu|^4$.
  We take into consideration this shift in $\lambda$ at $\Lambda = M$ when solving the RGEs,
\begin{align}
\lambda(\Lambda)\to \lambda(\Lambda)-\frac{5}{32\pi^2}|Y_\nu|^4.
\label{shift}
\end{align}

\begin{figure}[h]
\centering
\includegraphics[width=0.49\textwidth]{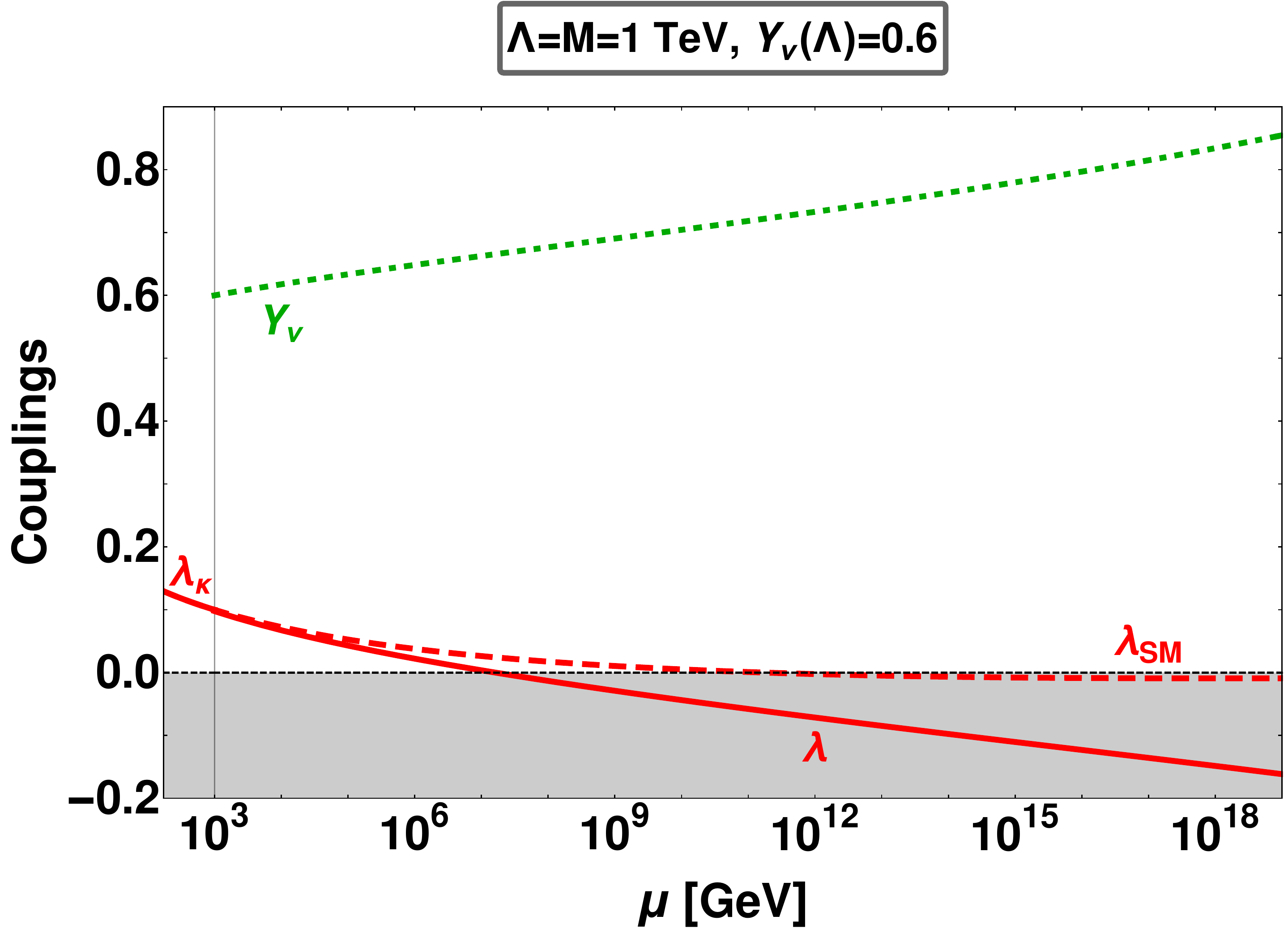}
\includegraphics[width=0.49\textwidth]{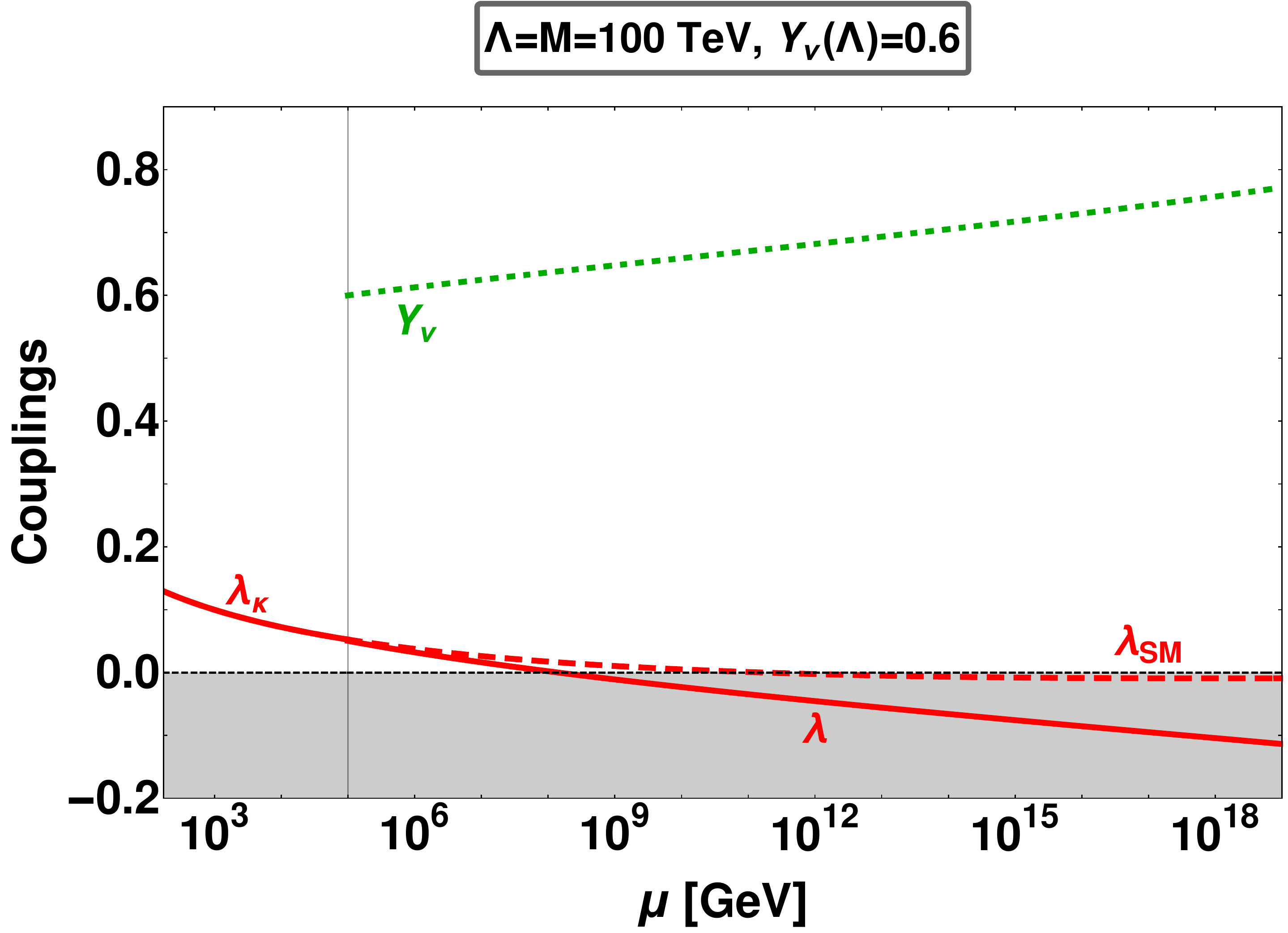}
\caption{\footnotesize{Evolution of the Higgs quartic self-coupling $\lambda$~(solid-red) and Yukawa coupling $Y_\nu$~(dotted-green) within the minimal (3,1,1) inverse seesaw scheme.
 $\lambda_\kappa$ is the quartic coupling in the effective theory with the Weinberg operator. 
 For comparison, we also plot the running of $\lambda_{\text{SM}}$, the SM quartic coupling, indicated by the dashed-red line.} }
\label{RG-311-Running}
\end{figure}

  Having set up our basic scheme, let us start by looking at the impact of the Yukawa coupling $Y_\nu$ on the stability of the Higgs vacuum.
  As already discussed, in the Standard Model, the running of the Higgs quartic coupling $\lambda_{\text{SM}}$ is dominated by the top quark Yukawa coupling and becomes negative around energy scale $\sim 10^{10}$ GeV.
However, within the inverse seesaw, the Yukawa coupling $Y_\nu$ in Eq.\eqref{lag-inv-seesaw} can dominate the evolution of $\lambda$ above the threshold scale $\Lambda = M$, as seen in Fig.~\ref{RG-311-Running}.

In Fig.~\ref{RG-311-Running} we have shown the RG evolution of the relevant coupling parameters assuming the Yukawa coupling $Y_{\nu}=0.6$ at the threshold scale,
taken to be $\Lambda = M = 10^3$ GeV (left panel) and $10^5$ GeV (right panel). 
  We see that $\lambda$ becomes negative at around energy scales $3.27\times 10^{7}$ GeV and $3.16\times 10^{8}$ GeV for the threshold scale $\Lambda =10^3$ GeV and $10^5$ GeV, respectively.
By comparing this with the running of the \sm Higgs quartic coupling $\lambda_{\text{SM}}$ (red dashed), one sees how the Higgs vacuum stability problem becomes more acute in the inverse seesaw model.
  This was expected, since the new fermions tend to destabilize the Higgs vacuum, as illustrated in Fig.~\ref{Yukawa negative effect}.
  It should also be noted that in the effective theory regime 
  the evolution of the quartic coupling $\lambda_\kappa$ almost coincides with that of $\lambda_{\text{SM}}$, due to the negligible effect of the Weinberg operator on its running.
Finally, note that all couplings in Fig.~\ref{RG-311-Running} remain within the perturbative region up to Planck scale.
  %


\subsection*{Consistency Restrictions}
\label{sec:cons-restr}

We now turn to the issue of the general self-consistency of the inverse seesaw mechanism.
  In order to ensure a perturbative and mathematically consistent model, the tree-level couplings must satisfy certain conditions,
  e.g. all of them should have a perturbative value, and the potential should be bounded from below. 
  However, once we take into account the quantum corrections, these conditions also get corrected. In this section we analyze these modified conditions in more detail. \\[-.2cm]

\begin{figure}[h]
\centering
\includegraphics[width=0.49\textwidth]{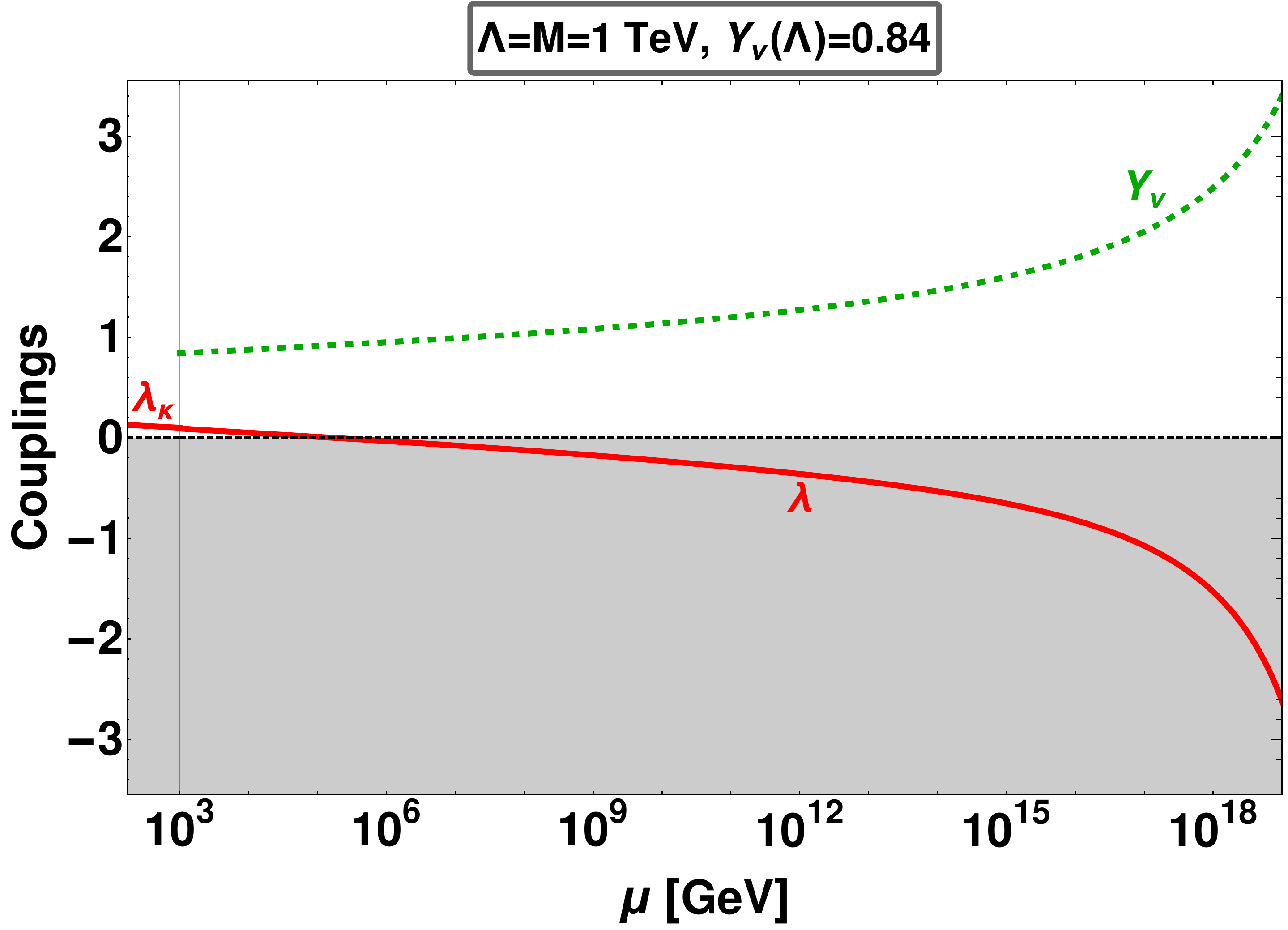}
\includegraphics[width=0.49\textwidth]{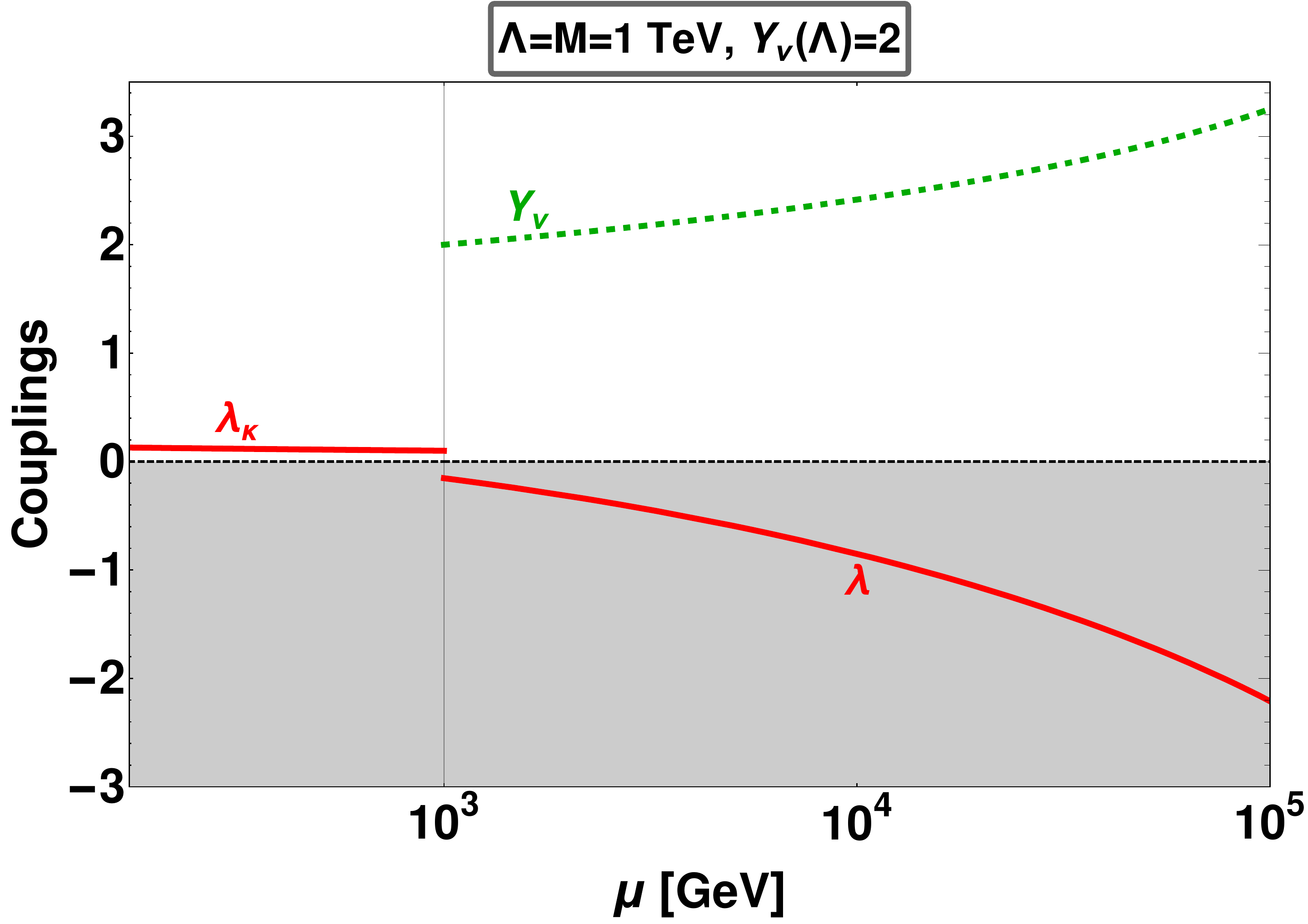}
\caption{\footnotesize{
%
    Perturbativity limits on the Yukawa coupling $Y_\nu$. The left panel requires $Y_\nu < \sqrt{4\pi}$ up to the Planck scale, so that only RG evolution is relevant.
    The right panel demands $Y_\nu < \sqrt{4\pi}$ only up to 100 TeV. In this case $Y_\nu$ is large enough that threshold effects make $\lambda$ negative even before running.
    In both cases the vacuum is unstable, i.e. $\lambda < 0$, before $Y_\nu$ reaches the perturbative limit, see text for details.
  }}
\label{fig:Yukawa-Bound}
\end{figure}

We start by examining the restrictions coming from perturbativity at tree-level, which require $|Y_\nu| < \sqrt{4 \pi}$.
  The RG evolution of $Y_\nu$ increases its value with increasing scale.
  Fig.~\ref{fig:Yukawa-Bound} shows the evolution of $Y_\nu$ and $\lambda$.
 From the left panel of Fig.~\ref{fig:Yukawa-Bound} one sees that demanding that $|Y_\nu| < \sqrt{4 \pi}$ up to the Planck scale implies that $|Y_\nu|\lsim 0.8$ at the threshold scale $ \Lambda = 10^3$ GeV.
 However, as one can see from Fig.~\ref{fig:Yukawa-Bound}, the Higgs quartic coupling $\lambda$ becomes negative much before the Planck scale.
 Therefore, demanding pertubativity of $Y_\nu$ all the way up to the Planck scale does not ensure full consistency of the scalar potential. 
%
 If one demands perturbativity only till, say, 100 TeV, as shown in right panel of Fig.~\ref{fig:Yukawa-Bound}, one finds that the pertubativity limit on $Y_\nu$ is relaxed to $|Y_\nu|\lsim 2$
 at the threshold scale $ \Lambda = 10^3$ GeV.
 Such large $Y_\nu$ values lead to large threshold corrections for $\lambda$ -- the negative jump shown in the right panel -- making it negative even before turning on its RG evolution. \\[-.2cm]

This highlights the importance of taking into account the threshold corrections for $\lambda$.
From Fig.~\ref{fig:Yukawa-Bound} one sees that a large $Y_\nu$ value can lead to an unbounded potential already at the threshold scale, even before RG evolution.
Taking the Yukawa coupling $Y_\nu(\Lambda)=1.58$ at $\Lambda=10^3$ GeV makes $\lambda(\Lambda)=0$ due to threshold corrections.
RG running will further decrease $\lambda$ above the threshold scale, making the vacuum unstable.
It is clear that threshold corrections are crucial when considering large Yukawa couplings and that a true limit on $Y_\nu$ requires one to take into account both RG evolution as well as the threshold
corrections it induces on the quartic coupling $\lambda$.
\begin{figure}[h]
\centering
\includegraphics[width=0.49\textwidth]{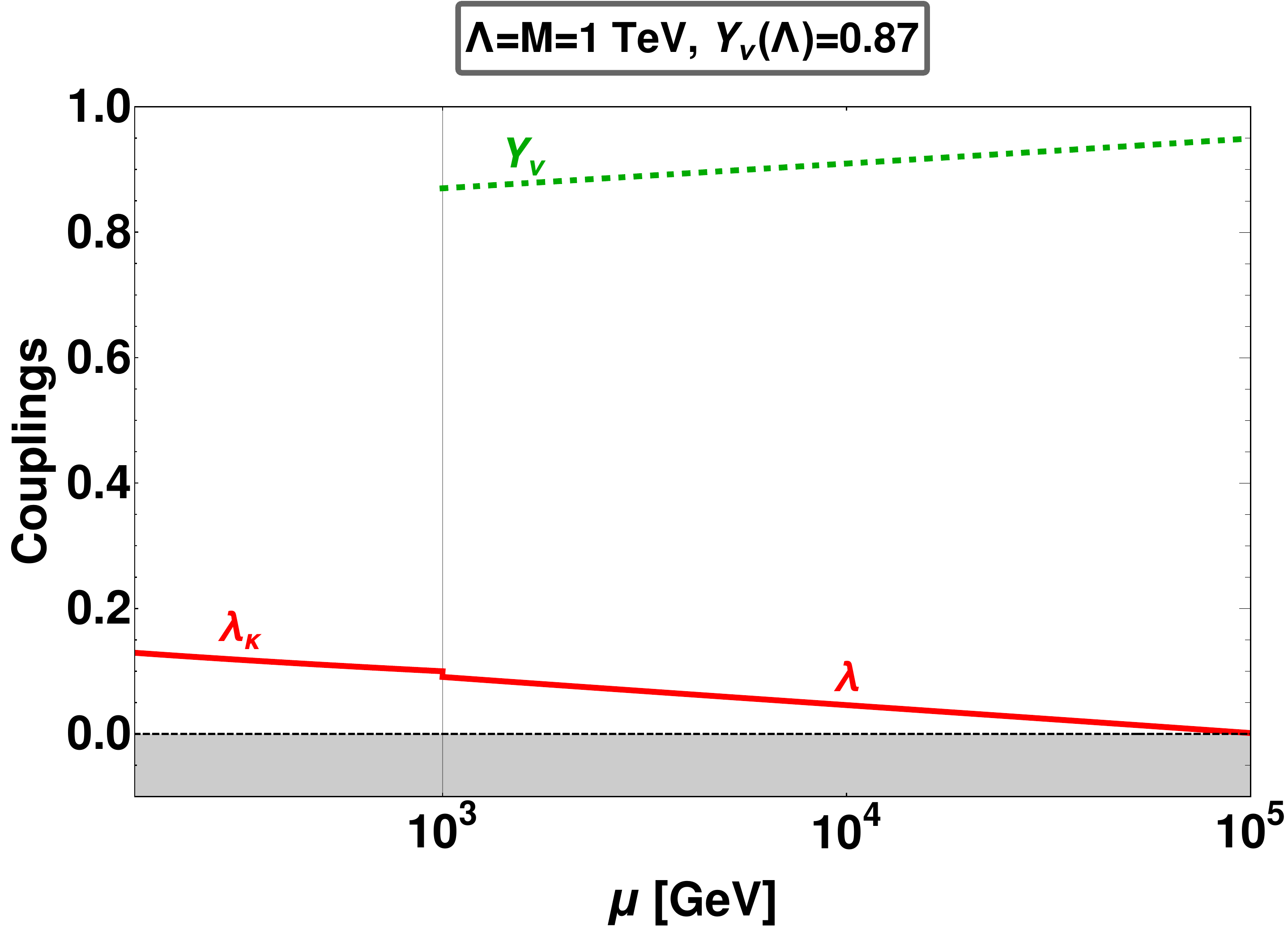}
\includegraphics[width=0.49\textwidth]{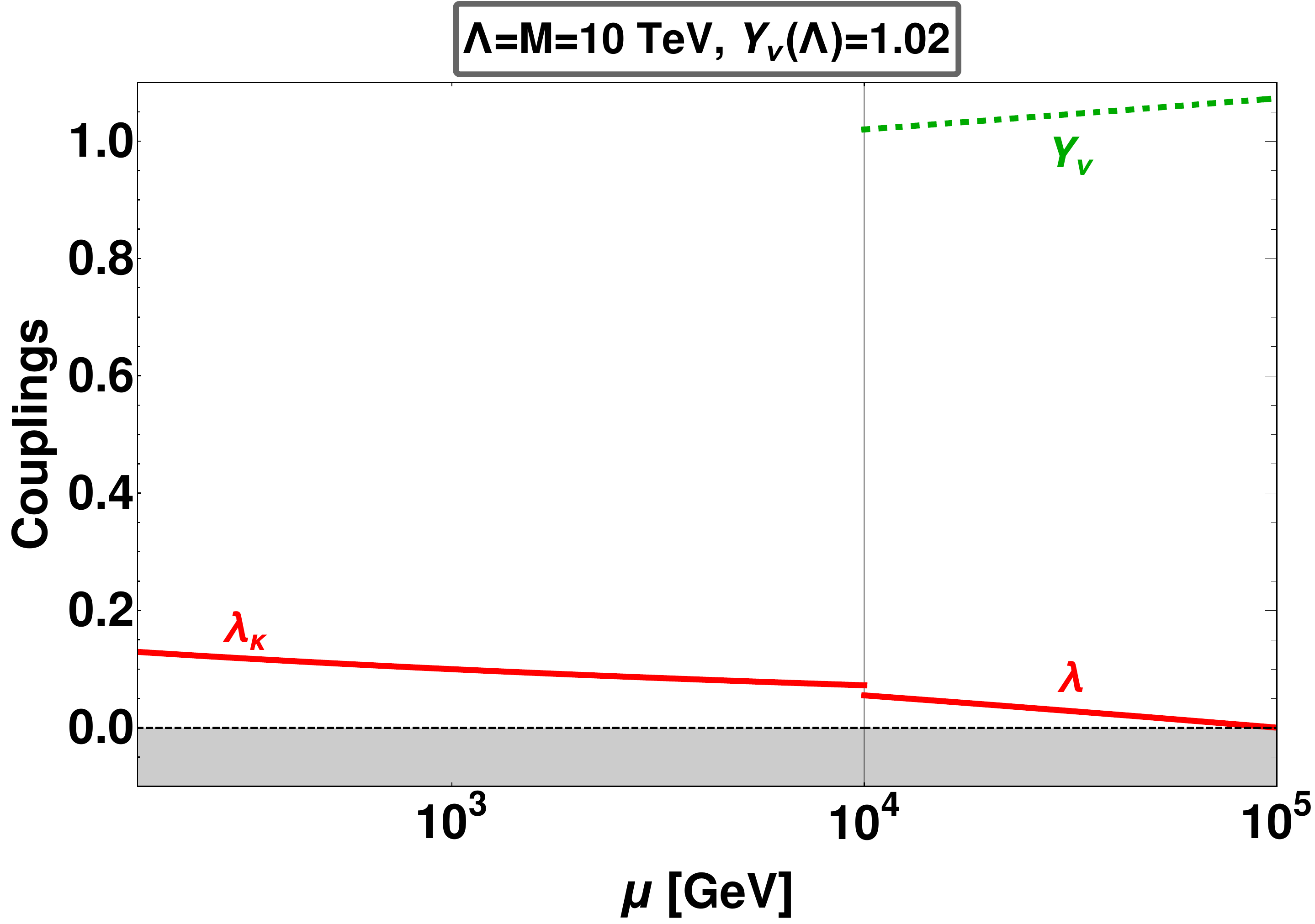}
\caption{\footnotesize{Limiting $Y_\nu$ by demanding $Y_\nu$ to remain perturbative and $\lambda$ to remain positive up to 100 TeV.
Left (right) panel correspond to threshold scales $\Lambda=1$ TeV ($\Lambda=10$ TeV). See text for details.}}
\label{fig:Boundedness100TeV}
\end{figure}

  As an example, in Fig.~\ref{fig:Boundedness100TeV} we show the result of demanding that neither $Y_\nu$ goes non-perturbative, nor $\lambda$ goes negative up to 100 TeV.
  To quantify the implications of this demand, we have taken two threshold scales, $\Lambda=10^3$GeV (left panel), and $\Lambda=10^4$GeV (right panel), respectively.
  With this combined requirement we obtain the limit $Y_\nu \lsim 0.87$ (left panel) and $Y_\nu\lsim 1.02$ (right panel).
  This illustrates that the limit on $Y_\nu$ also depends on the choices of threshold scale, for higher threshold scales the limit on $Y_\nu$ gets relaxed.

\section{The Majoron Completion of the inverse Seesaw}
\label{sec:major-compl-inverse}

In the previous section we saw that the addition of new fermions to the \sm in order to mediate neutrino mass generation via the inverse seesaw mechanism~\cite{Mohapatra:1986bd}
has a destabilizing effect on the Higgs vacuum. \black
  This problem can be potentially cured if there are other particles in the theory providing a ``positive'' contribution to the RGEs governing the evolution of the Higgs quartic coupling.
  A well-motivated way to do this is to assume the dynamical version of the inverse seesaw mechanism~\cite{GonzalezGarcia:1988rw}.

Building up on the work of Ref.~\cite{Bonilla:2015kna} here we focus on low-scale generation of neutrino mass through the inverse seesaw mechanism with spontaneous lepton number violation.
Lepton number is promoted to a spontaneously broken symmetry within the minimal $SU(3)_c\otimes SU(2)_L\otimes U(1)_Y$ gauge framework.
To achieve this, in addition to the Standard Model singlets $\nu^c$ and $S$, we now add a complex scalar singlet $\sigma$ carrying two units of lepton number.
  Lepton number symmetry is then spontaneously broken by the vacuum expectation value of $\sigma$. The relevant Lagrangian is given by
  \begin{align}
-\mathcal{L} = \sum_{i,j}^{3} Y_{\nu}^{ij} L_{i} \tilde{\Phi} \nu^c_{j} + M^{ij} \nu^c_{i}  S_{j} + Y_{S}^{ij} \sigma S_{i} S_{j} + \text{H.c.} 
\end{align}
After the electroweak and lepton number symmetry breaking the neutrino mass matrix has the following form 
\begin{align}
\mathcal{M}_{\nu}=
 \begin{pmatrix}
  0 & m_D  & 0 \\
  m_D^T & 0 & M \\
  0 &  M^T  &  \mu_S  \\
 \end{pmatrix}
\end{align}
where $m_D = \frac{Y_\nu v_\Phi}{\sqrt{2}}$, $\mu_S=2\frac{Y_Sv_\sigma}{\sqrt{2}}$ with $\vev{\Phi} = \frac{v_\Phi}{\sqrt{2}}$ and $\vev{\sigma} = \frac{v_\sigma}{\sqrt{2}}$ being the vacuum expectation values (vevs) of the $\Phi$ and $\sigma$ fields respectively. 
Again, within the standard seesaw approximation, the effective neutrino mass is obtained as
\begin{align}
 m_{\nu}\simeq \frac{v_\Phi^{2}} {\sqrt{2}} Y_{\nu} M^{-1} Y_{S} v_{\sigma} (M^T)^{-1} Y_{\nu}^{T}
\end{align}

Light neutrino masses of $\mathcal{O} (0.1)$ eV, are generated for reasonable choices of $v_{\sigma}$ and $M$, small Yukawa couplings $Y_{S}$, and sizeable $Y_{\nu} \sim \mathcal{O} (1)$.\\[-.2cm]

Turning to the scalar sector, in the presence of the complex scalar singlet $\sigma$ and doublet $\Phi$, the most general potential driving electroweak and lepton number symmetry breaking is given by
\begin{align}
 V=-\mu_{\Phi}^{2}\Phi^{\dagger}\Phi-\mu_{\sigma}^{2}\sigma^{\dagger}\sigma+\lambda_\Phi (\Phi^{\dagger}\Phi)^{2}+\lambda_{\sigma}(\sigma^{\dagger}\sigma)^{2}+\lambda_{\Phi\sigma}(\Phi^{\dagger}\Phi)(\sigma^{\dagger}\sigma).
\end{align}
As already noted, in addition to the $SU(3)_c\otimes SU(2)_{L}\otimes U(1)_Y$ gauge invariance, $V(\Phi,\sigma)$ also has a global $U(1)$ lepton number symmetry.

This  potential is bounded from below if $\lambda_\sigma$, $\lambda_\Phi$ and $\lambda_{\Phi\sigma} + 2 \sqrt{\lambda_\sigma \lambda_\Phi}$ are all positive,
and has a minimum for non-zero vacuum expectation values of both $\Phi$ and $\sigma$ provided $\lambda_\Phi$, $\lambda_\sigma$ and $4\lambda_\Phi\lambda_\sigma - \lambda_{\Phi\sigma}^{2}$ are all positive.
After the breaking of electroweak and lepton number symmetries, we end up with a physical Goldstone boson, the Majoron $J$~\cite{Chikashige:1980ui, Schechter:1981cv}, which is a pure gauge singlet.
After symmetry breaking one has, in the unitary gauge, 
\begin{align}
\Phi \to \frac{1}{\sqrt{2}}
 \begin{pmatrix}
 0 \\
 v_\Phi+h' \\
 \end{pmatrix}
,\hspace{1cm}
\sigma \to  \frac{v_{\sigma} + \sigma' + i J}{\sqrt{2}}.
\end{align}
The CP even fields $h'$ and $\sigma'$ will mix, so the mass matrix for neutral scalar $M_{ns}$ is given by
\begin{align}
 M_{ns}^2=
 \begin{pmatrix}
  2\lambda_\Phi v_\Phi^2 & \lambda_{\Phi\sigma}v_\Phi v_\sigma \\
  \lambda_{\Phi\sigma}v_\Phi v_\sigma & 2\lambda_\sigma v_{\sigma}^2 \\
 \end{pmatrix}
 \label{mass matrix for majoron case}
\end{align}
We can diagonalise the above mass matrix to obtain the mass eigenstates $(h \, \, H)^T$ through the rotation matrix $O_R$ as
\begin{align}
 \begin{pmatrix}
  h \\
  H  \\
 \end{pmatrix}
=O_R
\begin{pmatrix}
h' \\
 \sigma' \\
\end{pmatrix}
\equiv
\begin{pmatrix}
 \text{cos}~\alpha & \text{sin}~\alpha \\
 -\text{sin}~\alpha & \text{cos}~\alpha \\
\end{pmatrix}
\begin{pmatrix}
 h' \\
 \sigma' \\
\end{pmatrix},
\label{eq:rot}
\end{align}
Here $\alpha$ is the CP-even scalar mixing angle, and its range of allowed values is constrained by LHC data~\cite{Bonilla:2015uwa,Aad:2014iia}. The rotation matrix satisfies 
\begin{align}
 O_R M_{ns}^2 O_R^T=\text{diag}(m_h^2,m_H^2)
 \label{eq:diagonalisation-majoron-case}
\end{align}
where the masses $m_h, m_H$ of the scalars $h, H$ respectively, are given by 
\begin{align}
m_{h}^{2}=\lambda_\Phi v_\Phi^{2}+\lambda_{\sigma} v_{\sigma}^{2}-\sqrt{(\lambda_\Phi v_\Phi^{2}-\lambda_{\sigma} v_{\sigma}^{2})^{2}+(\lambda_{\Phi\sigma}v v_{\sigma})^{2}}
\label{eq:lig-scalar-mass} \\
m_{H}^{2}=\lambda_\Phi v_\Phi^{2}+\lambda_{\sigma} v_{\sigma}^{2}+\sqrt{(\lambda_\Phi v_\Phi^{2}-\lambda_{\sigma} v_{\sigma}^{2})^{2}+(\lambda_{\Phi\sigma}v v_{\sigma})^{2}}
\label{eq:hev-scalar-mass}
\end{align}
The lighter of these two mass eigenstates $h$ is identified with the $125$ GeV scalar discovered at the LHC~\cite{Aad:2012tfa,Chatrchyan:2012ufa}.

We can use Eqs.~\eqref{eq:lig-scalar-mass} and \eqref{eq:hev-scalar-mass} along with \eqref{mass matrix for majoron case}-\eqref{eq:rot} to solve for the parameters $\lambda_\Phi$, $\lambda_\sigma$ and $\lambda_{\Phi\sigma}$
in terms of physical quantitites i.e. masses $m_h^2$, $m_H^2$ and the mixing angle $\alpha$ as  
\begin{align}
 \lambda_\Phi&=\frac{m_h^2\cos^2\alpha+m_H^2\sin^2\alpha}{2v_\Phi^2},\\
 \lambda_\sigma&=\frac{m_h^2\sin^2\alpha+m_H^2\cos^2\alpha}{2v_\sigma^2},\\
 \lambda_{\Phi\sigma}&=\frac{(m_h^2-m_H^2)\sin\alpha \cos\alpha}{v_\Phi v_\sigma}.
\label{lambda definition}
\end{align}

\section{Vacuum stability in inverse seesaw with majoron}
\label{sec:vacu-stab-inverse}

\begin{figure}[h]
\centering
\includegraphics[width=0.3\textwidth]{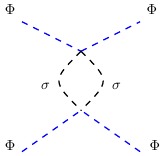}
\caption{\footnotesize{One-loop correction to the $\Phi$ quartic coupling due to its interaction with the singlet $\sigma$ that drives spontaneous lepton number violation in inverse seesaw models.
This diagram leads to a ``positive'' term in the RGE of the $\Phi$ quartic coupling, that can overcome the destabilizing effect of the fermions in Fig.~\ref{Yukawa negative effect}.} }
\label{Majoron contribution to lambda}
\end{figure}

In this section we will explore the consequences of spontaneous breaking of the lepton number symmetry on the stability of the electroweak vacuum.
Due to the presence of the scalar $\sigma$, the RGE of the $\Phi$ quartic coupling receives a new 1-loop contribution through the diagram shown in Fig.~\ref{Majoron contribution to lambda}.
This ``positive'' contribution plays a crucial role in counteracting the ``negative'' contribution coming from the extra fermions of the inverse seesaw model, see Fig.~\ref{Yukawa negative effect}.
%

Vacuum stability in this model can be studied in two different regimes namely i) $v_\sigma \gg v_\Phi$ and ii) $v_\sigma \approx \mathcal{O} (v_\Phi)$.
We start with the first possibility. As before, we focus on the missing partner (3, 1, 1) inverse seesaw, other possibilities will be taken up in Section~\ref{sec:comp-stand-miss}.
\subsection{Case I: $v_\sigma\gg v_\Phi$}

\begin{figure}[h]
\centering
\includegraphics[width=0.49\textwidth]{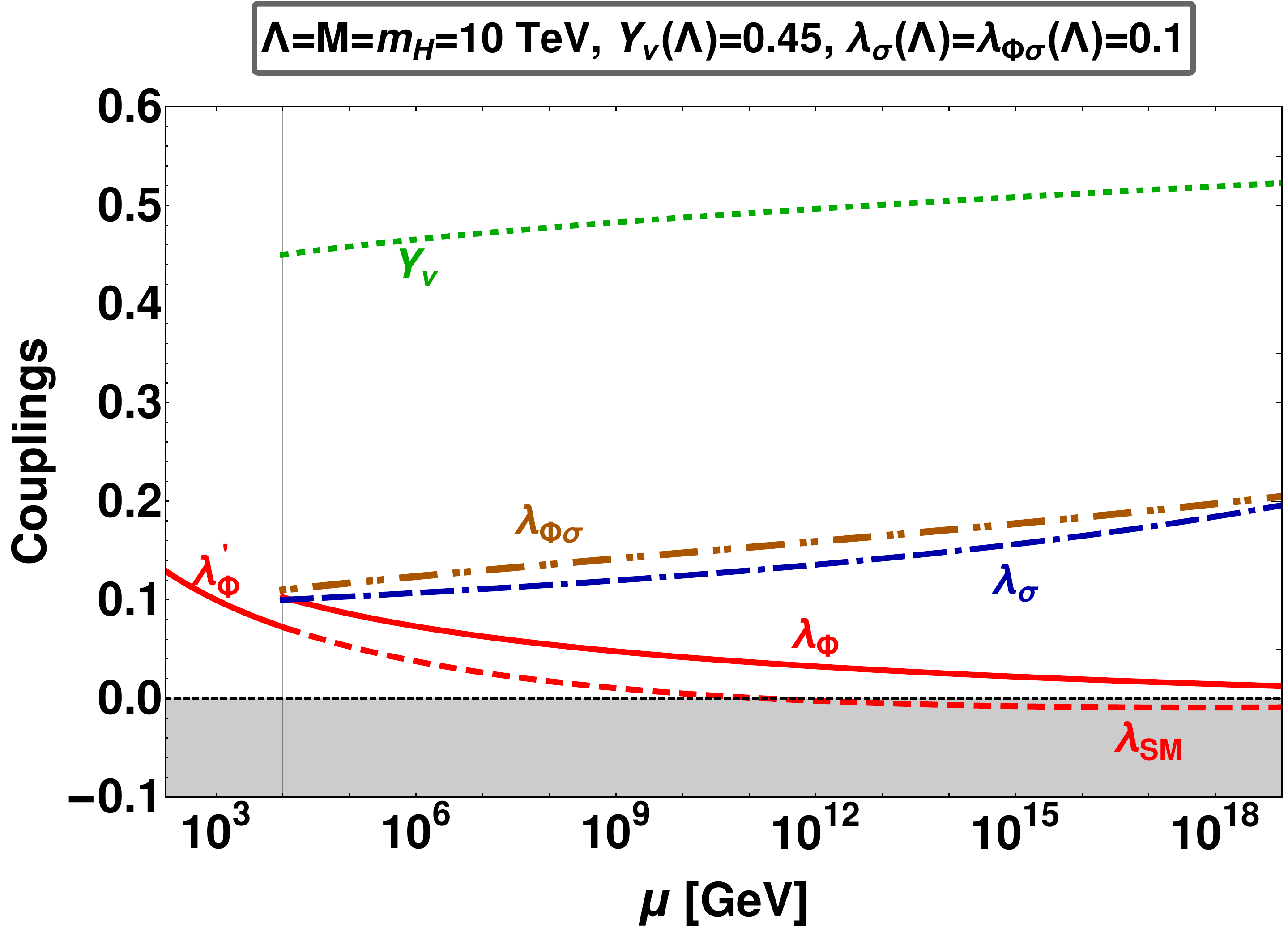}
\includegraphics[width=0.49\textwidth]{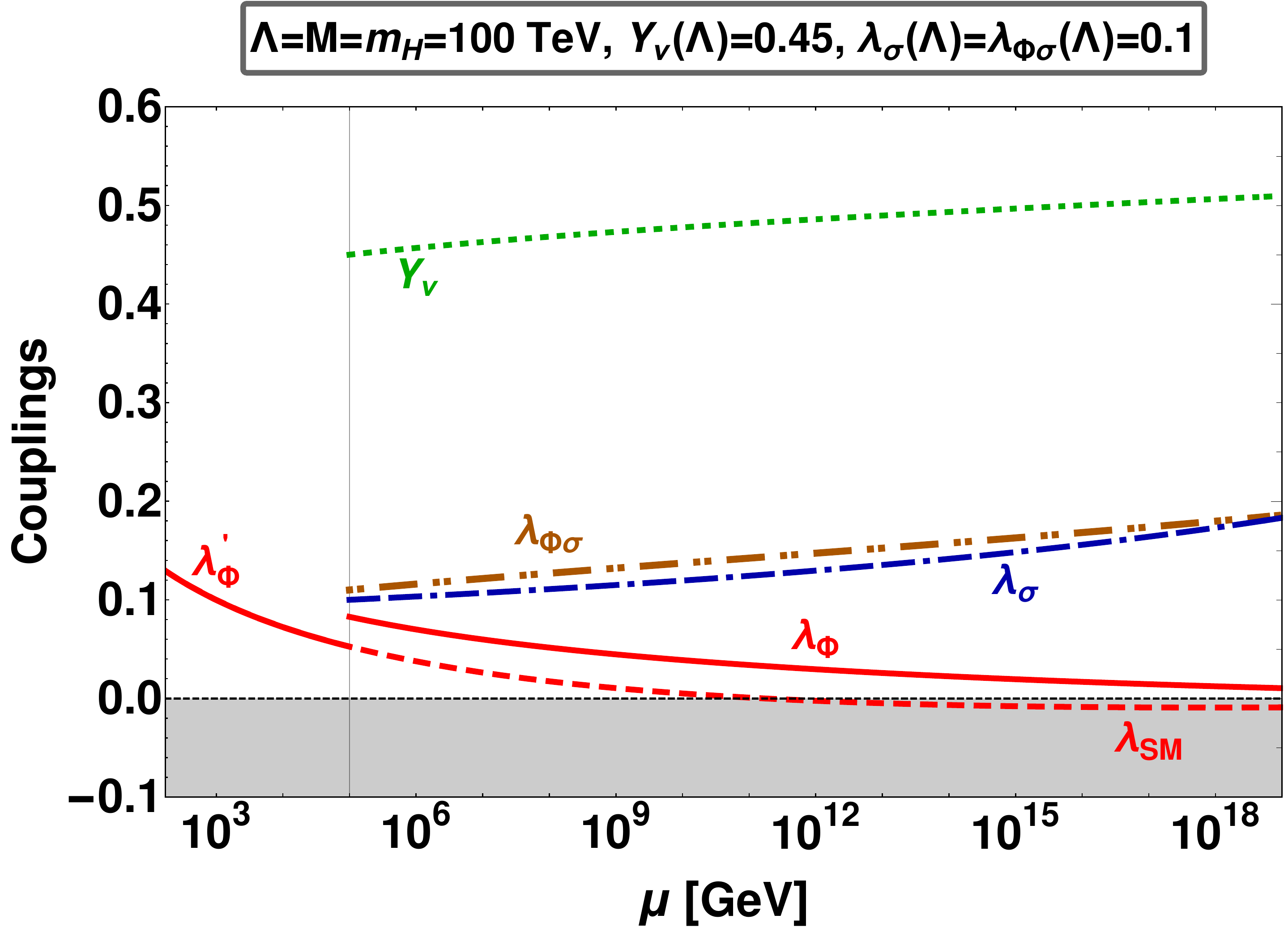}
\caption{\footnotesize{
    The RG evolution of the quartic couplings and right-handed neutrino Yukawa couplings within the Majoron extension of (3,1,1) inverse seesaw scheme.
    For comparison, we also show the evolution of $\lambda_{\text{SM}}$ (red-dashed). Here $\lambda_{\Phi}'\equiv \lambda_\kappa$ is the effective Higgs quartic coupling below threshold,
    see Eq.~\ref{threshold shift}.}
}
\label{RG-311MajoronThreshold-Running}
\end{figure}

In the limit $v_\sigma\gg v_\Phi$ the heavy CP-even Higgs boson $H$ almost decouples, with its mass $m_H$ given as $m_H\equiv  M_H \approx \sqrt{2\lambda_\sigma v_\sigma}$. 
Moreover, in this limit small neutrino masses require small $Y_S$, so the two heavy singlet fermions $\nu^c$ and $S$ form a quasi-Dirac pair with nearly degenerate mass $M$.
We assume, for simplicity of the analysis, that $M_H$ and $M$, have a common value, so that we deal with just one threshold scale $\Lambda = M = M_H$.
Below this scale we have an effective theory with the \sm structure, suplemented by the Weinberg operator for neutrino mass generation\footnote{Note that the majoron $J$ will also be present in this effective theory. Even though massless or fairly light, it will pratically decouple from the Higgs boson, and will not affect vacuum stability.}.
  Thus, below the threshold scale, we need to integrate out $\sqrt{2}\text{Re}(\sigma)$ at tree-level~\cite{EliasMiro:2012ay}. 
  As a result, at the scale $\Lambda$, there is a tree-level threshold correction which induces a shift in the Higgs quartic coupling,
  $\delta\lambda=\frac{\lambda_{\Phi\sigma}^2}{4\lambda_{\sigma}}$. This will lead to the following effective Higgs potential below the threshold scale $\Lambda$
\begin{align}
V_{\text{eff}}=\lambda_\Phi'\Big(\Phi^{\dagger}\Phi-\frac{v^2}{2}\Big)^2,
\end{align}
where the effective Higgs quartic coupling $\lambda_\Phi'$ below the threshold scale is defined as
\begin{align}
\lambda_\Phi'\equiv\lambda_\kappa=\lambda_\Phi-\frac{\lambda_{\Phi\sigma}^2}{4\lambda_\sigma}.
\label{threshold shift}
\end{align}
Here $\lambda_\kappa$ is the effective quartic coupling for the case of explicit lepton number breaking, see Section~\ref{sec:higgs-vacu-stab}.
The evolution of the Higgs quartic coupling $\lambda'_\Phi$ in the effective theory  is shown in Fig.~\ref{RG-311MajoronThreshold-Running}.
  One can appreciate the jump in the value of the Higgs quartic coupling due to threshold corrections. 
  Since only the dimension-five Weinberg operator runs below the scale $\Lambda$, the RG evolution of $\lambda_\text{SM}$ is essentially the same as that of $\lambda'_\Phi$.
  Both are very close to the RG running of $\lambda_\kappa$ of the effective theory with explicit lepton number breaking.
  Moreover, at tree-level the numerical value of $\lambda_\kappa(M_Z)$ and $\lambda_{\text{SM}}(M_Z)$ is the same, since in both cases one must reproduce the 125 GeV Higgs mass.

 Moving on to the full theory at the threshold scale $\Lambda = M$, the first thing to note is the impact of threshold corrections, Eq.~\ref{threshold shift}.
  They lead to a positive shift in value of the Higgs quartic coupling above the threshold scale $\Lambda = M$, enhancing the chances of keeping $\lambda_\Phi$ positive~\cite{Mandal:2019ndp}.
 Furthermore, to understand the evolution of $\lambda_\Phi$ in the full theory above the scale $\Lambda = M$ one must perform the RG evolution of all parameters.
 Above the scale $\Lambda$ one needs to include $\beta_{\lambda_{\Phi\sigma}}$, $\beta_{\lambda_\sigma}$ and evolve the quartic coupling $\lambda_\Phi$ using the full RGEs with the matching condition
 Eq.~\ref{threshold shift} at $\Lambda$.
 In Appendix.~\ref{app:inverse seesaw with majoron}, we give the two-loop RGEs of the full theory. 

   In Fig.~\ref{RG-311MajoronThreshold-Running} we show the evolution of various couplings in the majoron inverse seesaw model for given benchmark points.
  We have taken the threshold scale as $M = M_H = 10$ TeV and $M = M_H = 100$~TeV for the left and right panels, respectively.  
  For the sake of comparison, the initial values of other parameters have been kept the same in both panels. The Yukawa coupling has been fixed at $Y_\nu=0.45$.
  We have taken $\lambda_\sigma,\lambda_{\Phi\sigma}=0.1$ at the scale $ \Lambda$. 
  The positive shift in the evolution of $\lambda$ at the threshold scale is coming from the matching condition given in Eq.~\ref{threshold shift}.
  Notice that below threshold the running of $\lambda_\Phi'$ and $\lambda_\text{SM}$ almost coincide with each other, due to the tiny effective Weinberg operator. 
  Finally, since $Y_{S}$ has been taken to be very small, it has no direct impact on vacuum stability. 

  In summary, it is clear from Fig.~\ref{RG-311MajoronThreshold-Running} that the dynamical variant of the inverse seesaw mechanism can be free from the Higgs vaccum instability problem. 
  This is possible thanks to the positive contribution of the scalar $\sigma$ both to the threshold corrections, as well as to the RG evolution of the Higgs quartic coupling.
  These effects are enough to counteract the negative contribution of the new fermions present in inverse seesaw model, even for sizeable Yukawa couplings $Y_\nu \sim \mathcal{O} (1)$.
These could lead to a plethora of new phenomena~\cite{Dittmar:1989yg,Abreu:1996pa,Acciarri:1999qj,Cai:2017mow,Bernabeu:1987gr,Branco:1989bn,Rius:1989gk,Deppisch:2004fa,Deppisch:2005zm,Valle:1987gv,Nunokawa:1996tg,Antusch:2006vwa,Miranda:2016ptb,Escrihuela:2015wra}.
  Thus, in contrast to the case of inverse seesaw with explicitly broken lepton number, the dynamical variants can have a completely stable Higgs vacuum.

\subsection{Case II: $v_\sigma=\mathcal{O}(v_\Phi)$}

  In this case, the mass of the heavy scalar $m_H$ is of the order of the electroweak scale. 
Hence we can neglect the small range between $M_Z$ and $m_H$, starting instead with Eq.~\ref{lambda definition}, which already includes the threshold effect of Eq.~\ref{threshold shift}. 
Thus in this case only the fermions are integrated out at the threshold scale $\Lambda = M$, while all the scalars remain in the resulting theory below threshold.
Thus the scalar couplings evolve over a larger range, and have better chance of curing the Higgs vacuum instability problem.
Needless to say that, as before, the Higgs vaccum instability can be avoided if the mixed quartic $\lambda_{\Phi \sigma}$ is sufficiently large, $\mathcal{O}(0.1)$.
This in turn implies a sizeable mixing $\alpha \sim \mathcal{O}(0.1)$ between the two CP-even Higgs bosons. 

\begin{figure}[h]
\centering
\includegraphics[width=0.49\textwidth]{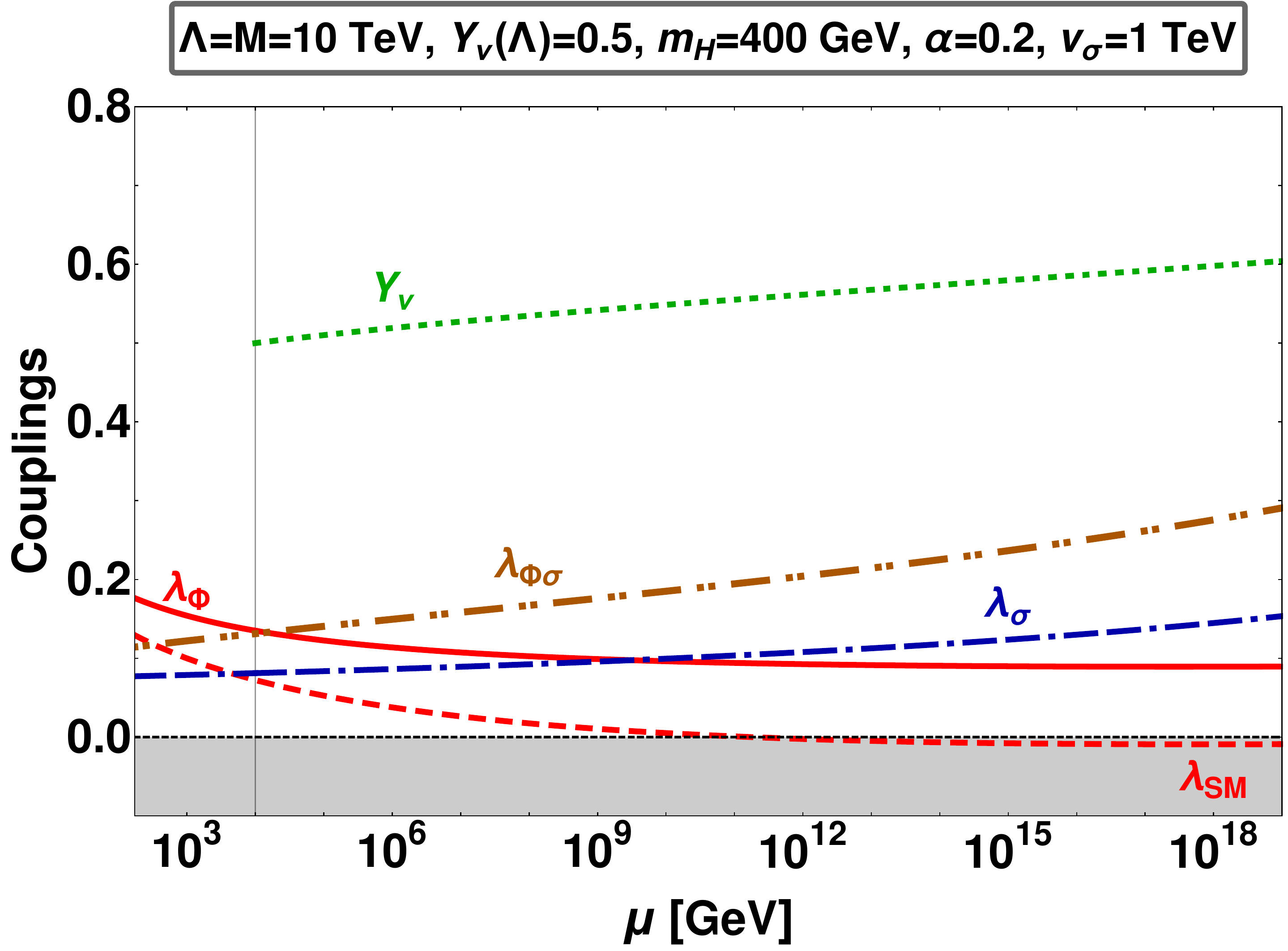}
\includegraphics[width=0.49\textwidth]{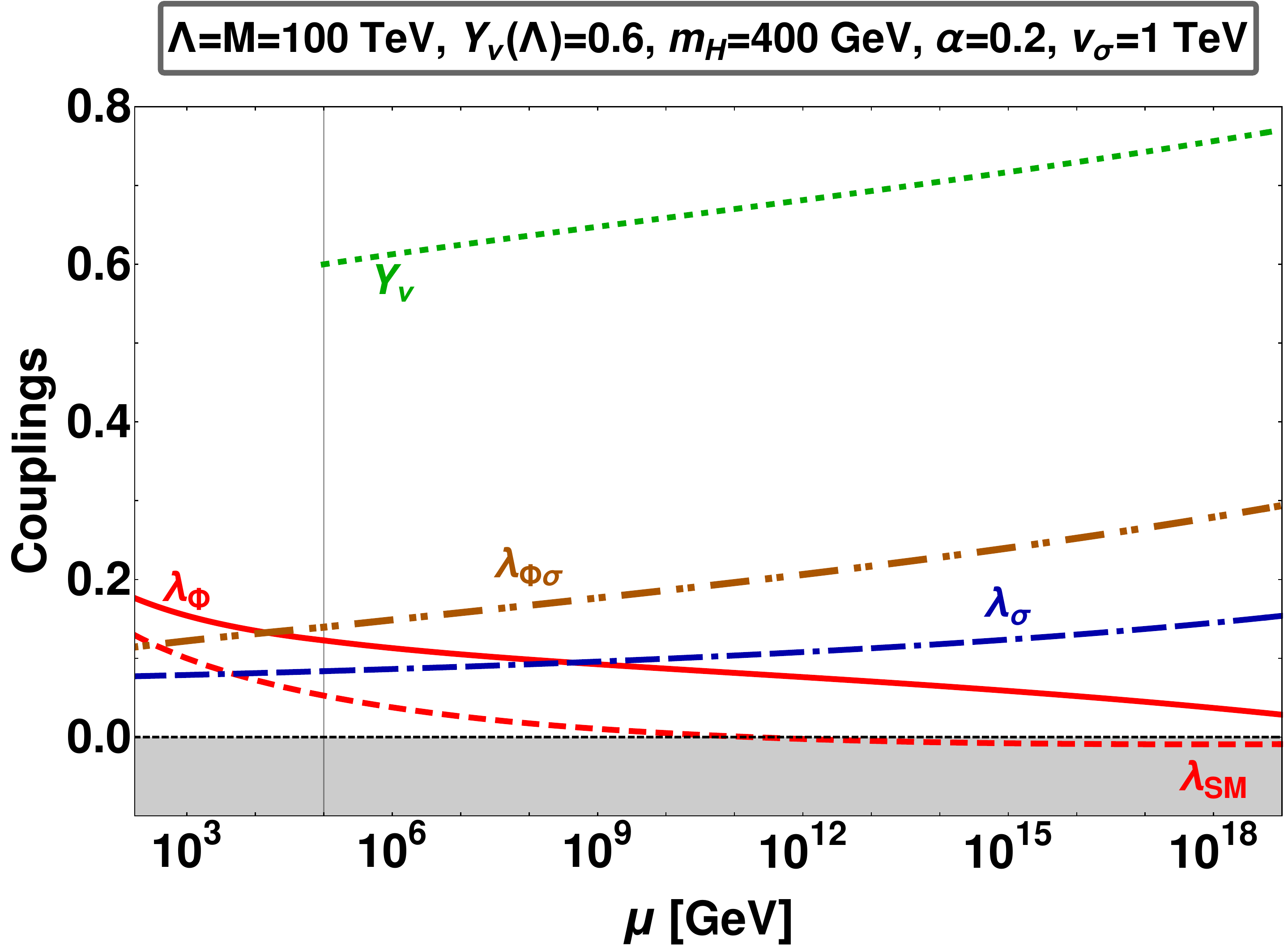}
\includegraphics[width=0.49\textwidth]{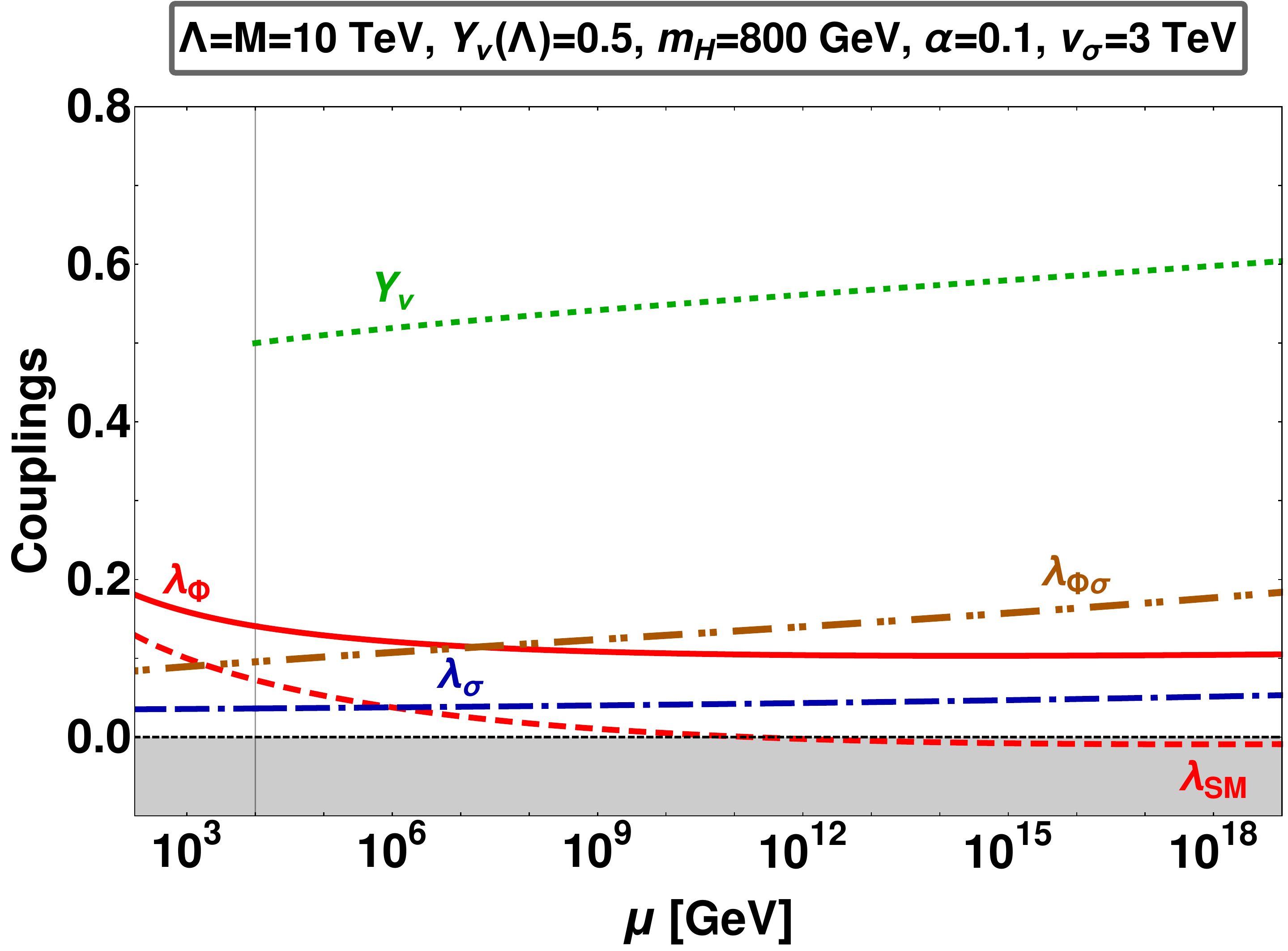}
\includegraphics[width=0.49\textwidth]{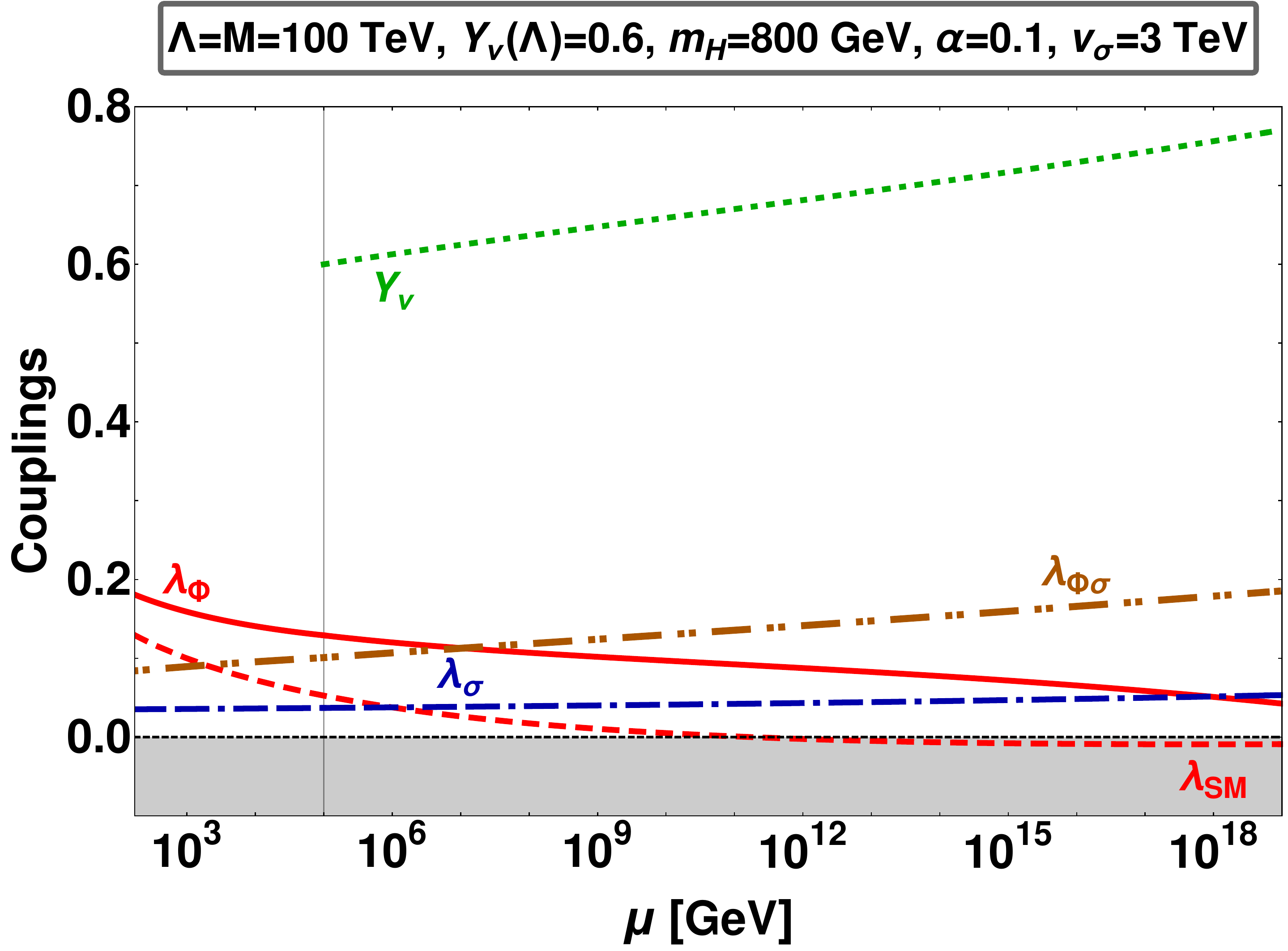}
\caption{\footnotesize{
      Evolution of the quartic couplings and right-handed neutrino Yukawas within the Majoron extension of the missing partner (3,1,1) inverse seesaw scheme.
       For comparison, the evolution of $\lambda_{\text{SM}}$ is shown in the red dashed curve.
      Here only the fermion singlets are integrated out at the threshold scale $\Lambda = M$, all scalars are part of the effective theory below threshold, taken as the weak scale. }}
\label{RG-311Majoron-Running}
\end{figure}

The evolution of the couplings in this case is shown in Fig.~\ref{RG-311Majoron-Running}.
In these plots, we have fixed the singlet neutrino scale $\Lambda = 10$ TeV in the left panel, and 100 TeV in the right panel.
In contrast to the scalar couplings, the Yukawa coupling $Y_\nu$ starts running only above threshold. 
Notice that for relatively  large mediator scale, the allowed value of $Y_\nu$ will also be  large as there is not enough range, in terms of RGEs evolution, to sizeably alter the $Y_\nu$.
We found that for large Yukawa couplings, $Y_\nu\geq 0.7$~(0.8) for threshold scale $\Lambda = 10$~TeV~(100 TeV), respectively, we get either unstable vacuum or non-perturbative dynamics.

Moreover, as shown in Fig.~\ref{RG-311Majoron-Running}, we can have positive $\lambda_\Phi$, $\lambda_\sigma$ and $\lambda_{\Phi\sigma}$ all the way up to the Planck scale, even for sizeable Yukawa couplings.
  We found that for small $m_H$ the required mixing angle is relatively large, in contrast to the large $m_H$ case.
  For small $\alpha$ or $m_H$ the potential becomes unbounded from below at high energies.
  In other words, experimental limits on $\alpha$, e.g. coming from the LHC~\cite{Bonilla:2015uwa,Aad:2014iia}, can be used to place a lower limit on the mass $m_H$.
  In Sec.~\ref{sec:invisible} we illustrate the interplay between the vacuum stability restrictions and the constraints on the invisible width of the Higgs boson that follow from current
  LHC experiments.
  There we also note that in order to prevent the existence of Landau poles in the running parameters, the lepton number breaking scale $v_\sigma$ should not be too small.
%


\section{Comparing sequential and missing partner inverse seesaw}
\label{sec:comp-stand-miss}

For simplicity we have so far only analyzed the explicit and dynamical lepton number breaking within the simplest (3,1,1) missing partner inverse seesaw mechanism. 
We now compare the stability properties of this minimal construction with those of (3,2,2) and (3,3,3) inverse seesaw mechanisms. 

\subsection{Sequential versus missing partner seesaw: electroweak vacuum stability}
\label{sec:sequ-vers-miss-1}

\begin{figure}[h]
\centering
\includegraphics[width=0.59\textwidth]{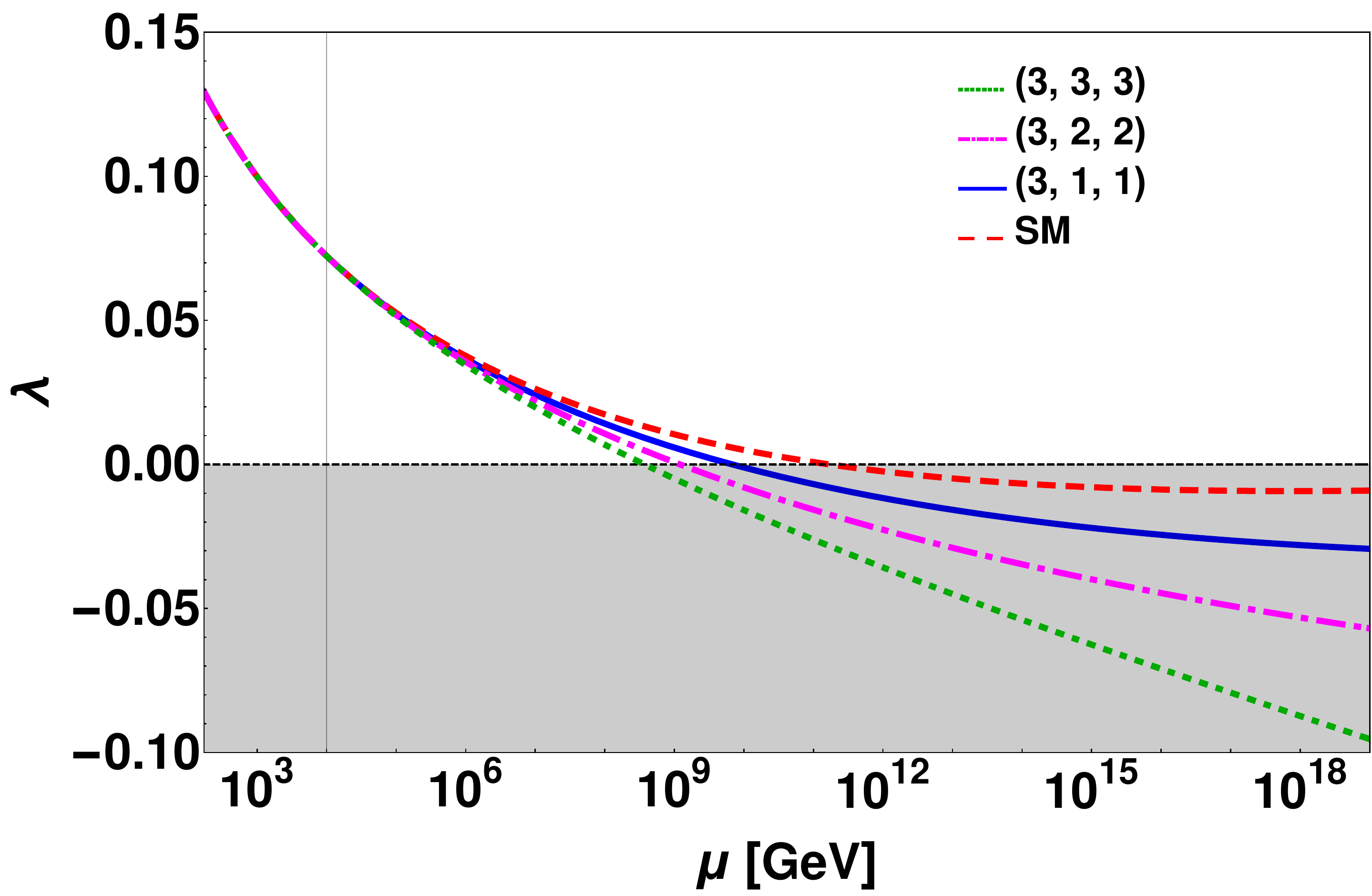}
\caption{\footnotesize{
     Comparing the evolution of the quartic Higgs self-coupling $\lambda$ in the Standard Model (dashed, red) with various inverse-seesaw extensions with explicit lepton number violation:
    (3,1,1) denoted in solid (blue), (3,2,2) dot-dashed (magenta) and (3,3,3) dotted (green), see text for details.
%
}}
\label{Three generation case}
\end{figure}

As already mentioned, the problem of Higgs vacuum stability only gets worse with the addition of extra fermions.
  This fact is clearly illustrated in Fig.~\ref{Three generation case} where we compare the RG evolution of the Higgs quartic coupling $\lambda$ within the \sm (dashed, red)
  with the $(3,n,n)$ inverse seesaw completions, with $n=1$ (solid, blue), $n=2$ (dot-dash, magenta) and $n=3$ (dot, green). \\[-.2cm]

  In Fig.~\ref{Three generation case} we have taken the initial Yukawa coupling values in such a way as to facilitate a proper comparison of the different cases.
  To do this for (3,1,1) case, we have fixed the Yukawa coupling $|Y_\nu|=0.4$.
  For (3,2,2) and (3,3,3) case, we have taken the diagonal entries of the $Y_\nu$ matrix to be $Y_\nu^{ii}=0.4$, while all off-diagonal ones, $Y_{\nu}^{ij}$ for $i\ne j$, were neglected in the RGEs.
  Clearly one sees how $(3,n,n)$ inverse seesaw scenarios with $n>1$ have worse Higgs vacuum stability properties than the $n=1$ case.\\[-.2cm]
\begin{figure}[h]
\centering
\includegraphics[width=0.49\textwidth]{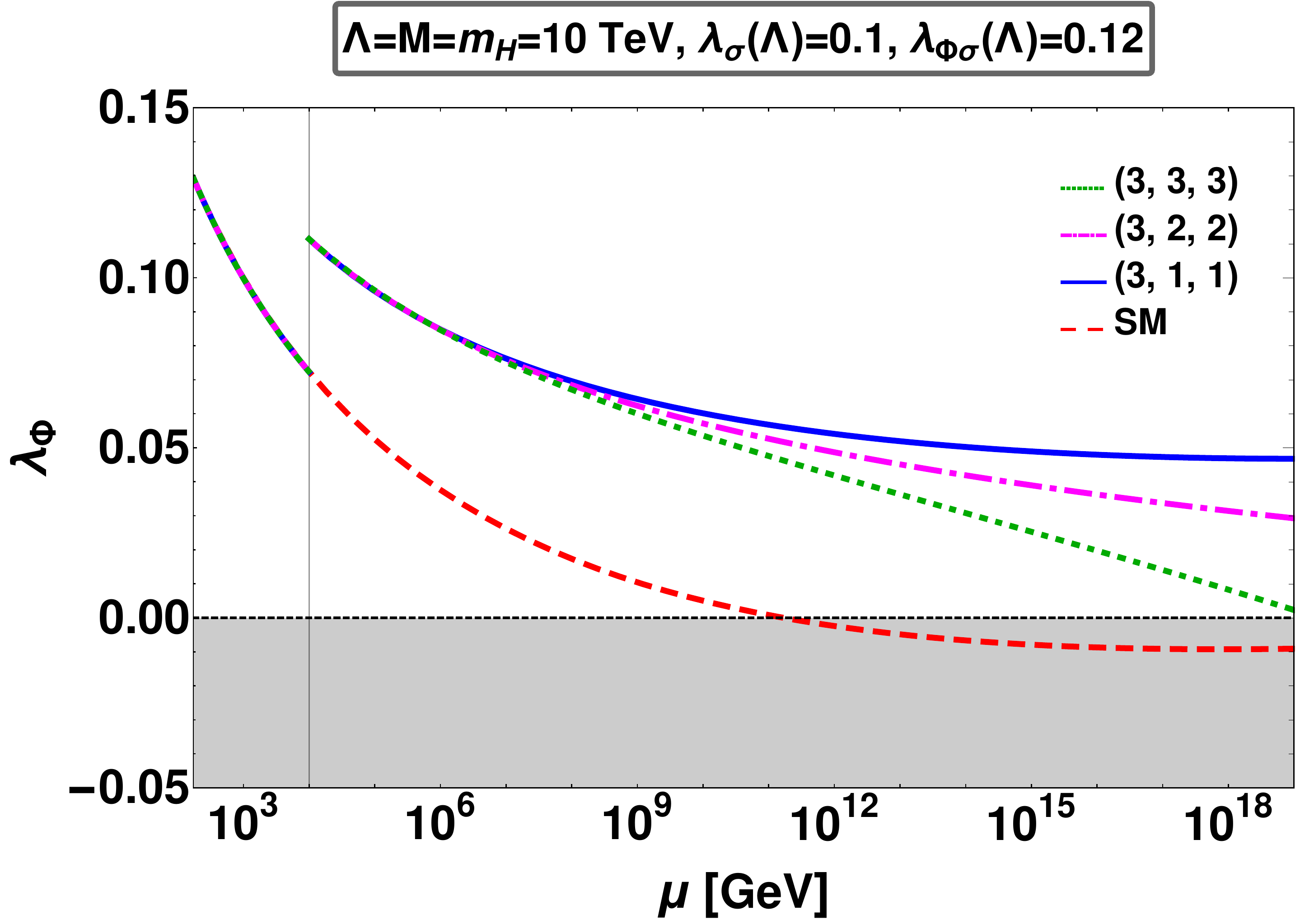}
\includegraphics[width=0.49\textwidth]{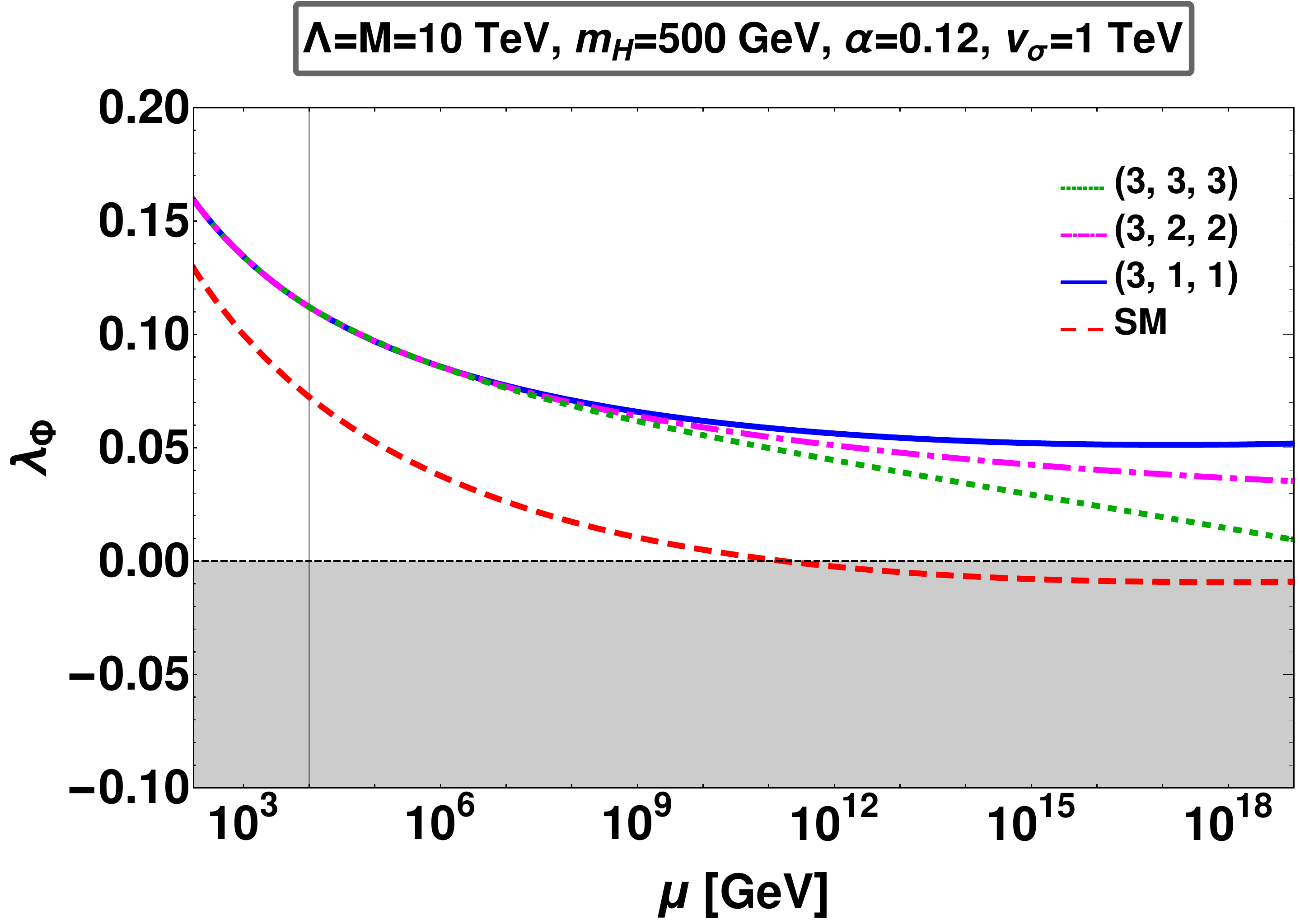}
\caption{\footnotesize{
    Comparing the evolution of the quartic Higgs self-coupling $\lambda$ in the Standard Model (red dashed) with the majoron inverse seesaw mechanism:
   the minimal (3,1,1) is denoted in solid (blue), (3,2,2) is dot-dashed (magenta) and (3,3,3) is dotted (green). See text for details. 
}}
\label{Three generation Majoron case with 1 TeV}
\end{figure}
%

In Fig.~\ref{Three generation Majoron case with 1 TeV}, we display our vacuum stability results for the majoron inverse seesaw models. 
One can compare the \sm case (dashed, red) with the (3,1,1) (solid, blue), (3,2,2) (dot-dash, magenta) and (3,3,3) (dot, green) majoron inverse seesaw schemes. 
As before, to ensure a consistent comparison, we have taken the Yukawa coupling $|Y_\nu|=0.4$ for (3,1,1) case, while
for the (3,2,2) and (3,3,3) cases, we have taken $Y_\nu^{ii}=0.4$ and neglected off-diagonal $Y_{\nu}^{ij}$. 
In the left panel we have taken the case of $\Lambda = M = m_H  = 10$ TeV.
Below threshold we have integrated out the fields $\sqrt{2}\text{Re}(\sigma)$, $\nu^c$ and $S$ and included the threshold effects. 
This leads to the jump in the quartic coupling seen in the figure. 
In contrast, for the right panel, we have fixed $v_\sigma=1$ TeV and $m_H=500$ GeV.
In this case the scalars are not integrated out and the quartic coupling runs smoothly from electroweak scale till Planck scale.\\[-.2cm] 

Fig.~\ref{Three generation Majoron case with 1 TeV} clearly illustrates that even for $n\geq 2$,  we can have a stable electroweak vacuum for adequate choices of $\alpha$ and $m_H$.
Indeed, even in the higher (3,2,2) and (3,3,3) majoron inverse seesaw, the positive contribution from the new scalar is enough to overcome the negative contribution
from the new fermions of the inverse seesaw. 
In short, the Higgs vacuum can be kept stable all the way up to the Planck scale even for appreciable Yukawa coupling $Y_\nu$. 

\black
\subsection{Sequential versus missing partner seesaw: brief phenomenological discussion }
\label{sec:sequ-vers-miss-2}

Here we note that neither the explicit nor the dynamical variant of the minimal (3,1,1) inverse seesaw mechanism is phenomenologically realistic.
The reason is that (3,1,1) leads to only one massive neutrino (lying say, at the atmospheric scale), hence inconsistent with oscillation data~\cite{deSalas:2020pgw}.
This minimal scheme is simply the inverse seesaw embedding of the minimum ``missing partner'' (3,1) seesaw mechanism of Sec.III in Ref.~\cite{Schechter:1980gr}.
This lack of the solar neutrino mass splitting can be avoided by the presence of a complementary radiative mechanism.
To implement such ``radiative completion'' of the minimal scheme one would need to invoke new physics.
The latter could be associated, say, to the presence of a dark matter sector~\cite{Rojas:2018wym}.
This would provide an elegant theory with a tree-level atmospheric scale, and a radiatively-induced solar neutrino mass scale, very much analogous to the case of the bilinear breaking of R-parity
in supersymmetry~\cite{Diaz:1997xc,Hirsch:2000ef,Diaz:2003as}. \\[-.2cm] 

Alternatively, one can generate non-zero tree-level masses for two neutrinos by going directly to the (3,2,2) ``missing partner'' seesaw scheme.
Again, this would be the inverse-seesaw-analogue of the (3,2) seesaw mechanism in Ref.~\cite{Schechter:1980gr}.
Finally, the sequential (3,3,3) inverse seesaw mechanism will generate tree-level masses for all three light neutrinos.
Any of these would be totally consistent with neutrino oscillations~\footnote{Modulo, of course, explaining the detailed pattern of mixing angles indicated by the oscillation
  data~\cite{deSalas:2020pgw}. Such a challenging task would require a family symmetry, whose detailed nature is not yet fully understood.}.\\[-.2cm] 

Concerning neutrinoless double beta decay, here lies an important phenomenological difference between the ``missing partner'' and the ``sequential'' seesaw mechanism. 
In the missing partner seesaw there can be no cancellation amongst the individual light-neutrino amplitudes leading to the decay~\cite{Valle:2020wdf}~\footnote{This feature may also
  be implemented in some radiative schemes of scotogenic type, see e.g.~\cite{Reig:2018ztc,Leite:2019grf,Avila:2019hhv}.}.
As a result, there is a lower bound on the neutrinoless double beta decay rates that could be testable in the upcoming generation of searches. \\[-.2cm] 

There are other implications of low-scale seesaw schemes, such as our inverse-seesaw, that could be potentially testable in current or upcoming experiments.
  For example, the associated heavy neutrino mediators could be accessible at high energy experiments such as $e^+e^-$ collider~\cite{Dittmar:1989yg,Abreu:1996pa,Acciarri:1999qj,Cai:2017mow},
  with stringent bounds, e.g. from the Delphi and L3 collaborations~\cite{Abreu:1996pa,Acciarri:1999qj}.
 Likewise, they could produce interesting signatures at the LHC~\cite{Deppisch:2015qwa,Sirunyan:2018mtv}.
Moreover, these mediators would also induce lepton flavour and leptonic CP violation effects with potentially detectable rates,
unsuppressed by the small neutrino masses~\cite{Bernabeu:1987gr,Branco:1989bn,Rius:1989gk,Deppisch:2004fa,Deppisch:2005zm}. 
Finally, since the heavy singlet neutrinos would not take part in oscillations, these could reveal new features associated to unitarity violation in the lepton mixing matrix~\cite{Valle:1987gv,Nunokawa:1996tg,Antusch:2006vwa,Miranda:2016ptb,Escrihuela:2015wra}. 
A dedicated study would be required to scrutinize whether these signatures could be used to distinguish missing partner from sequential seesaw.

\section{Impact of invisible Higgs decay on the vacuum stability}
\label{sec:invisible}

As we saw above, vacuum stability is often threatened by the violation of the condition $\lambda_{\Phi}>0$.
  From the RGE running of $\lambda_{\Phi}$ in Eq.~\ref{lambdaphi} one sees that in order to overcome the destabilizing effect coming from
  fermions~($-6y_t^4$ and $-2\text{Tr}(Y_\nu Y_\nu^{\dagger}Y_\nu Y_\nu^{\dagger})$), one needs a relatively large mixed quartic coupling $\lambda_{\Phi\sigma}$.
  This in turn translates into a large mixing angle $\alpha$ between the $CP$-even neutral Higgs bosons $h$ and $H$.
  We see from Eq.~\ref{lambda definition} that large $\lambda_{\Phi\sigma}$ implies smaller mixing angle  $|\sin\alpha|$ for larger $m_H$ and vice-versa.
  Within dynamical low-scale seesaw schemes with $v_\sigma\sim\mathcal{O}(\text{TeV})$, relatively large mixing angle $|\sin\alpha|$ is expected.
  This is in potential conflict with the invisible Higgs decay constraints from LHC.
\begin{figure}[h]
\centering
\includegraphics[width=0.59\textwidth]{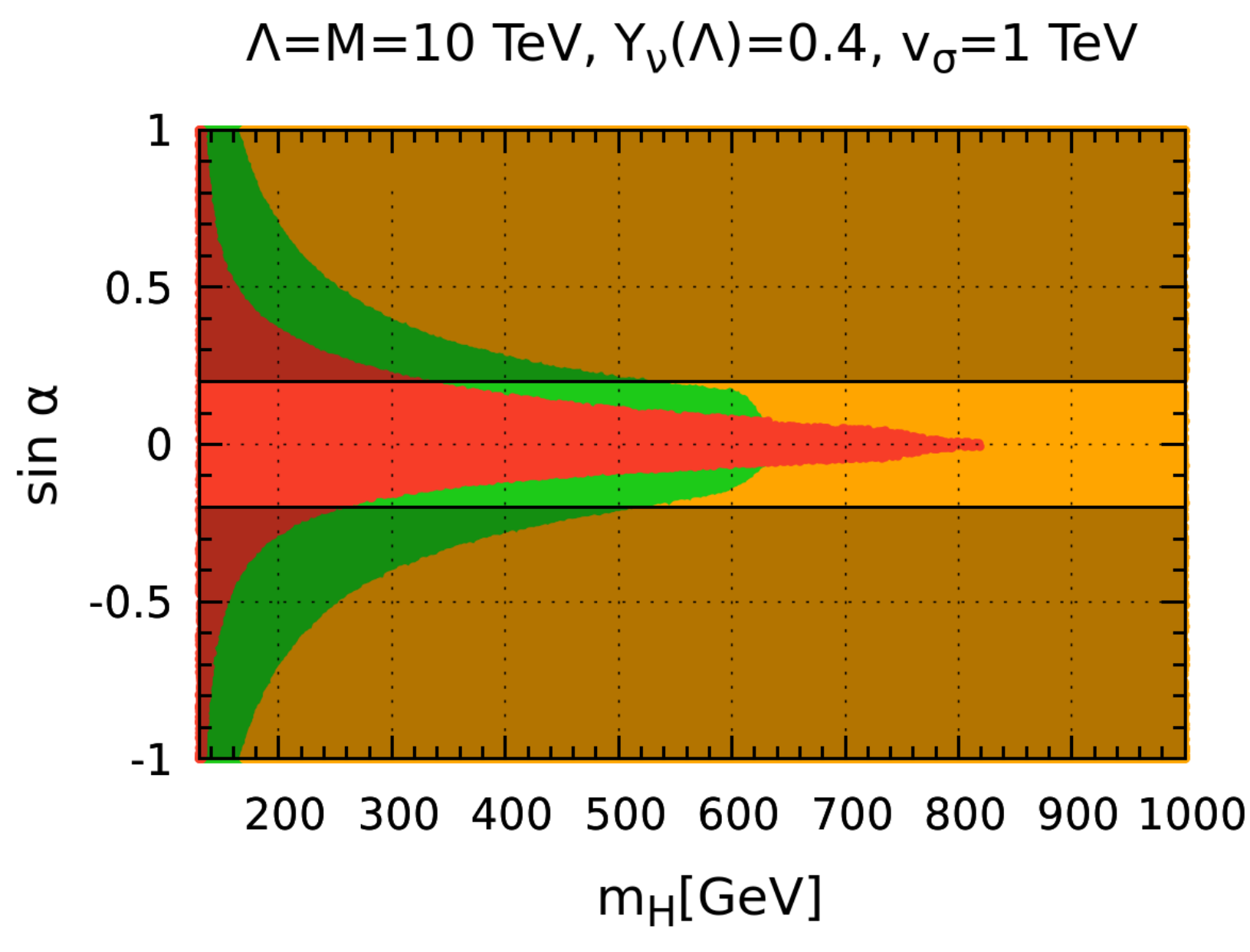}
\caption{\footnotesize{ Values of $m_H$ and mixing angle $\alpha$ leading to a stable potential (green), an unstable potential (red) and non-perturbative dynamics (orange).
 Here we take the (3,1,1) missing partner majoron inverse seesaw as the reference, with the heavy fermion threshold scale fixed as $\Lambda = 10$ TeV, Yukawa coupling $Y_\nu=0.4$ and $v_\sigma$ = 1 TeV.
 Within the green region all couplings are perturbative and the vacuum is stable up to the Planck scale. In the red region the potential becomes unbounded from below before the Planck scale.
 The orange region has nonperturbative couplings (including Landau poles) at energy scales below the Planck scale.
 The region outside the horizontal band delimited by the black lines is ruled out by the LHC constraints on invisible Higgs decays.  More details in text.}  }
\label{invisible-consistency}
\end{figure}

Indeed, it has long been noted that models with spontaneous violation of global symmetries such as lepton number at low scales $v_\sigma\sim\mathcal{O}(\text{TeV})$ lead to sizeable invisible Higgs decays, i.e. $h\to JJ$~\cite{Joshipura:1992hp} where $J$ is the Majoron.
  The existence of such invisible decays can be probed by the LHC experiments~\cite{Bonilla:2015uwa,Bonilla:2015jdf,Fontes:2019uld}. 
 The tightest bound on invisible Higgs boson decays comes from the CMS experiment at the LHC, $\text{BR}(h\to\text{Invisible})\leq 19\%$~\cite{Sirunyan:2018owy}.
  This upper limit on the invisible Higgs decay sets a tight constraint on $\lambda_{\Phi\sigma}$ or $|\sin\alpha|$ for $m_{H}>130$ GeV.
  For example with $v_\sigma = 1$ TeV one gets $|\sin\alpha|<0.2$ for $m_{H}>130$ GeV.\\[-.3cm]

  So far in all of our discussions we have chosen the mixing angle $|\sin\alpha|$ for fixed $v_\sigma$ and $m_H$ in such a way that one has consistency with the CMS constraint on invisible Higgs decay.
  However, the full parameter space of the model contains regions consistent with vacuum stability but disallowed by the invisible Higgs decay constraints.
  We illustrate this in Fig.~\ref{invisible-consistency} for the (3,1,1) missing partner seesaw with relatively large Yukawa coupling $Y_\nu=0.4$.
  Fig.~\ref{invisible-consistency} shows the values of $m_H$ and $\alpha$ for $v_\sigma$ = 1 TeV which lead to either stable/unstable potential or non-perturbative dynamics, as follows: 
 \begin{itemize}
   \item \textbf{Green Region:}
     In this region we have a stable vacuum all the way up to the Planck scale, with all the couplings within the perturbative regime. 
 In our numerical scan these conditions are implemented in following ways:  $0<\lambda_\Phi(\mu)<4\pi$, $0<\lambda_\sigma(\mu)< 4\pi$, $\lambda_{\Phi\sigma}(\mu)+2\sqrt{\lambda_\Phi(\mu)\lambda_\sigma(\mu)}>0$, $|\lambda_{\Phi\sigma} (\mu)|< 4\pi$ and $|Y_\nu(\mu)|< 4\pi$  where $\mu$ is the running mass scale. 
  \item \textbf{Red Region:}
    In this region the potential becomes unbounded from below at some high energy scale before the Planck scale.
    The potential is unbounded from below if any (or more) of the following conditions is realised:
    $\lambda_\Phi(\mu)\leq 0$, $\lambda_\sigma(\mu)\leq 0$ or $\lambda_{\Phi\sigma}(\mu)+2\sqrt{\lambda_\Phi(\mu)\lambda_\sigma(\mu)}\leq 0$. 
  \item \textbf{Orange Region:}
    Here one or more couplings become non-perturbative below the Planck scale. 
    This happens if any one of the following conditions holds: $|\lambda_\Phi(\mu)|\geq 4\pi$, $|\lambda_\sigma(\mu)|\geq 4\pi$, $|\lambda_{\Phi\sigma}(\mu)|\geq 4\pi$, $|Y_\nu(\mu)|\geq 4\pi$.
    Note that the possibility of Landau poles is also included inside the non-perturbative regions. 
\item \textbf{Collider constraint:}
    This is the region disallowed due to the LHC restriction on the Higgs invisible decay branching fraction which requires  $\text{BR}(h \to\text{Invisible}) \leq 19\%$~\cite{Sirunyan:2018owy}. 
   \end{itemize}
  From Fig.~\ref{invisible-consistency} one sees that for small $m_H$, the required mixing angle is large, in order to ensure a stable electroweak vacuum.
   This is in turn in conflict with the invisible Higgs decay constraints.
   As a result, one sees that these constraints are complementary to the vacuum consistency requirements of pertubativity and stability.
 Altogether, these can rule out a large part of the model parameter space.  \\ [-.3cm]
  
\begin{figure}[t]
\centering
\includegraphics[width=0.49\textwidth]{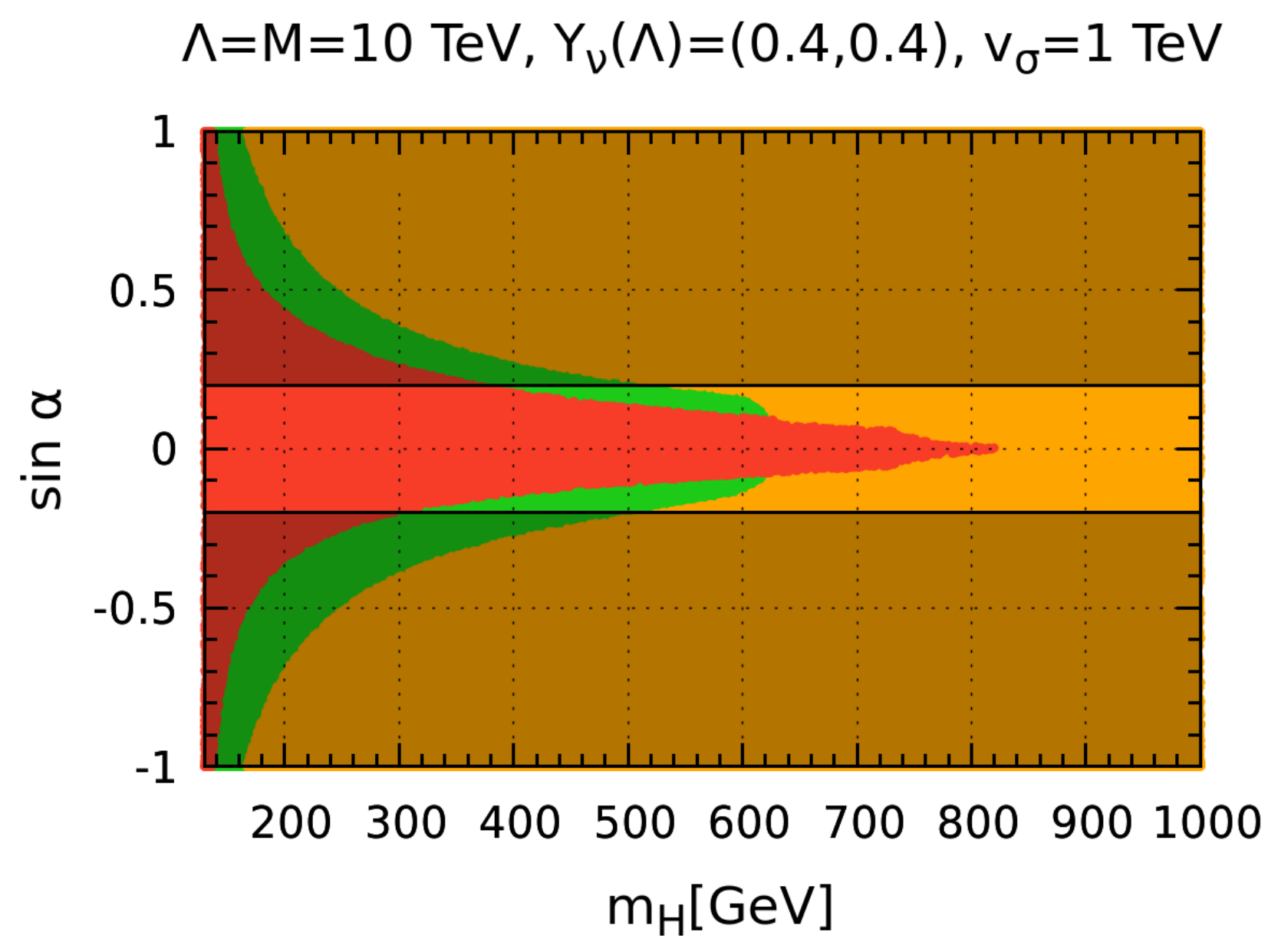}
\includegraphics[width=0.49\textwidth]{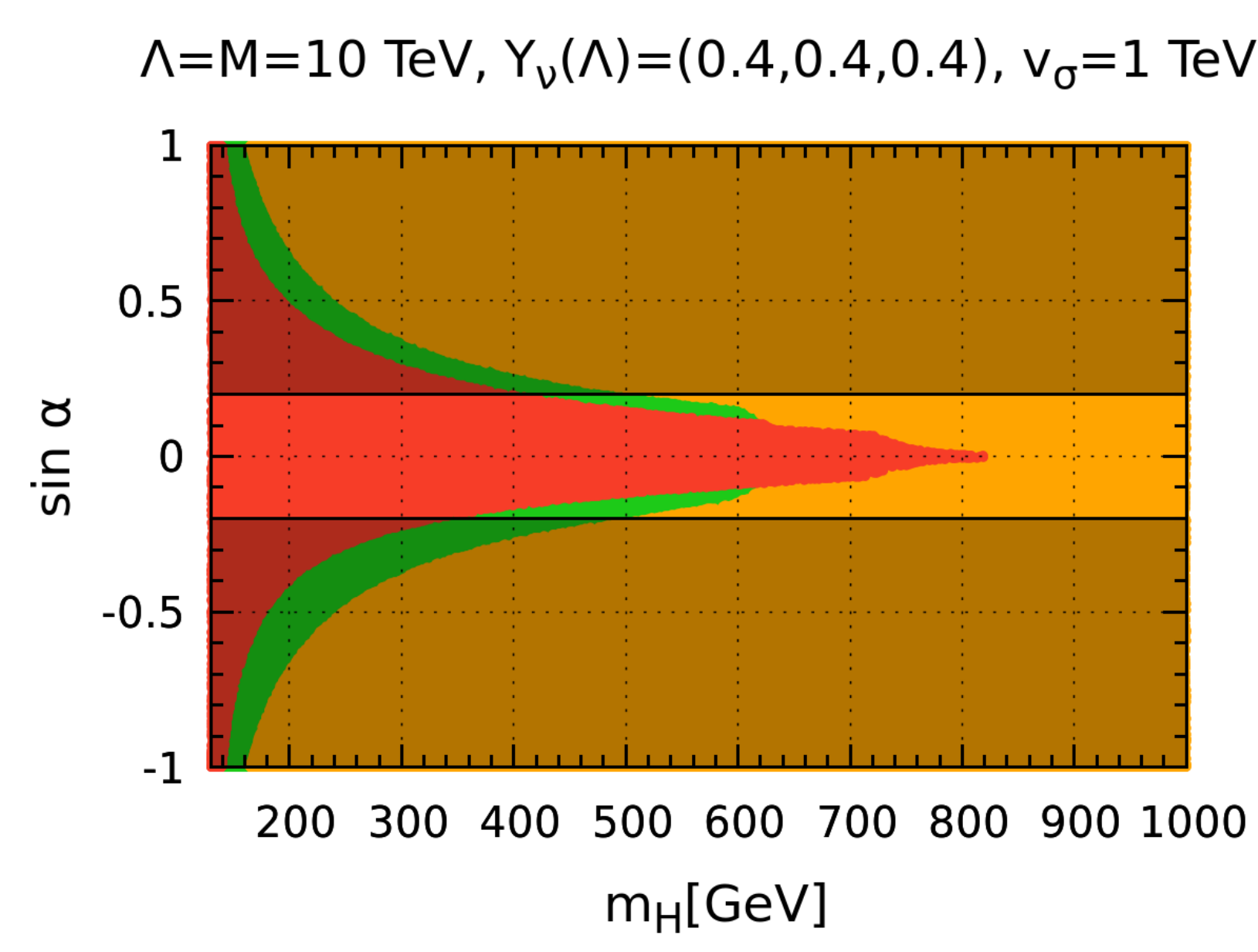}
\caption{\footnotesize{
    Vacuum consistency constraints of Fig.~\ref{invisible-consistency} for the case of (3,2,2)~(left) and (3,3,3)~(right) inverse majoron seesaw mechanism.
    The diagonal entries of the $Y_\nu$ matrix are fixed as $Y_\nu^{ii} = 0.4$. See text.   }  }
\label{invisible-consistency-2}
\end{figure}

 The above discussion refers to our (3,1,1) majoron inverse seesaw reference case, template for the scoto-seesaw mechanism~\cite{Rojas:2018wym}.
 One may now wonder how this discussion will change in the higher $(3,n,n)$; $n \geq 2$ inverse seesaw schemes which do not require a ``completion'' so as to generate the atmospheric scale. 
 In Fig.~\ref{invisible-consistency-2} we display the results for the (3,2,2) (left panel) and (3,3,3) (right panel) scenarios. 
 As expected, the undesired effect of additional fermions on the stability of the vacuum is clearly visible.
 Indeed, the unstable red regions in Fig.~\ref{invisible-consistency-2} are larger than in Fig.~\ref{invisible-consistency}.
 Likewise, the same effect is seen by comparing the left and right panels of Fig.~\ref{invisible-consistency-2}. 
 It is clear from Fig.~\ref{invisible-consistency} and Fig.~\ref{invisible-consistency-2} that the allowed parameter space consistent with stability and LHC constraints in the $(3,n,n)$ seesaw with
 $n\geq 2$ is more tightly restricted than in our reference $n=1$ case.
 However we note that, for moderate values of the Yukawa coupling, we still have parameter regions where electroweak breaking is consistent with the LHC measurements. 
 Although in the above we discussed (3,1,1), (3,2,2) and (3,3,3) cases separately, one should note that, in terms of RGE evolution, there is not much difference between them. 
 The corresponding RGEs (see appendix) are the same by replacing $|Y_\nu|^2$ by $\text{Tr}(Y_\nu^{\dagger}Y_\nu)$.
 Hence as long as one takes $|Y_\nu|^2 \approx \text{Tr}(Y_\nu^\dagger Y_\nu)$, the (3,1,1) and $(3, n, n)$ with $n\geq 2$ schemes are effectively the same.

Note that the restriction on the mixing angle gets stronger for lower values of $v_\sigma$ and weakens for higher values of $v_\sigma$, disappearing for high enough $v_\sigma$. 
  Therefore, the LHC measurements constitute a probe of the lepton number violation scale $v_\sigma$ associated to neutrino mass generation. 
  Moreover, note that here we have only considered the case when the lighter of the two CP even scalars is identified as the 125 GeV Higgs boson.
  \emph{A priori}, the possibility that the heavier CP even scalar is the 125 GeV Higgs boson should also be discussed.   
Finally, in the discussions of Fig.~\ref{invisible-consistency} and Fig.~\ref{invisible-consistency-2} we have required vacuum stability and perturbativity all the way up to the Planck scale. 
This will be an over-requirement, if there is other new physics at play. 
In that case one should require vacuum stability and perturbativity only up to a lower energy scale, say only up to 100 TeV, thus relaxing the resulting restrictions.
All of these issues require a dedicated study, that lies beyond the scope of the present work.


\section{Conclusions} 
\label{sec:Conclusions}
We have examined the consistency of electroweak symmetry breaking within the inverse seesaw mechanism.
We have derived the full two-loop renormalization group equations of the relevant parameters within inverse seesaw schemes,
examining both the simplest inverse seesaw with explicit violation of lepton number, as well as the majoron extension of inverse seesaw.
The addition of fermion singlets ($\nu^c$ and $S$) has a destabilizing effect on the running of the Higgs quartic coupling $\lambda$.
We found that for the inverse seesaw mechanism with sizeable Yukawa coupling $Y_\nu$ the quartic coupling $\lambda$ becomes negative much before the \sm instability scale $\sim 10^{10}$ GeV.
We have taken as our simplest benchmark neutrino model the ``incomplete'' (3,1,1) inverse seesaw scheme, as it has the ``best'' stability properties within this class of seesaw schemes.
We compared this reference case, in which only one oscillation scale is generated at tree-level, with the ``higher'' inverse seesaw constructions $(3,n,n)$ with $n=2,3$, in which other mass scales,
such as the atmospheric scale, also arise from the tree-level seesaw mechanism.
Our main results on the stability of the electroweak vacuum are summarized in Figs.~\ref{RG-311-Running}, \ref{RG-311MajoronThreshold-Running}, \ref{RG-311Majoron-Running}, \ref{Three generation case} and \ref{Three generation Majoron case with 1 TeV}.  
We showed how, in contrast to simplest inverse seesaw with explicit lepton number violation, the stability properties improve when this violation is spontaneous,
and there is a physical Nambu-Goldstone boson, the majoron.
The comparison with LHC restrictions is given in Figs.~\ref{invisible-consistency} and~\ref{invisible-consistency-2}.
We found that the LHC measurements constitute a probe of the lepton number violation scale $v_\sigma$ associated to neutrino mass generation. 
Its detailed study, however, needs further investigation. For example, we have assumed the lighter of the two CP even scalars to be the 125 GeV Higgs boson.
The alternative intriguing possibility should \emph{a priori} also be considered.   
We have also required vacuum stability and perturbativity all the way up to the Planck scale. This is clearly an over-requirement, in the presence of additional new physics.
The latter could be associated say, to dark matter or to the strong CP problem. 
In such case one should require vacuum stability and perturbativity only up to a lower intermediate energy scale, thus relaxing the restrictions we have obtained.
All of these issues require a dedicated study, that lies beyond the scope of the present work.

\black
\appendix
\label{Appendix}

\section{RGEs: Inverse seesaw}
\label{sec:rges:-inverse-seesaw}


In our work we have used the package SARAH~\cite{Staub:2015kfa} to perform the RG analysis. The $\beta$ function of a given parameter $c$ is given by,
\begin{align*}
 \mu\frac{dc}{d\mu}\equiv\beta_{c}=\frac{1}{16\pi^{2}}\beta_{c}^{(1)}+\frac{1}{(16\pi^{2})^{2}}\beta_{c}^{(2)} \, .
\end{align*}
where $\mu$ is the running scale and $\beta_{c}^{(1)}$, $\beta_{c}^{(2)}$ are the one-loop and two-loop RG corrections.
\subsection{Higgs quartic scalar self coupling}

The one-loop and two-loop RG corrections to the Higgs quartic self-coupling are given by
 \begin{align} 
& \beta_{\lambda}^{(1)}  =  
+\frac{27}{200} g_{1}^{4} +\frac{9}{20} g_{1}^{2} g_{2}^{2} +\frac{9}{8} g_{2}^{4} -\frac{9}{5} g_{1}^{2} \lambda -9 g_{2}^{2} \lambda +24 \lambda^{2}
 +12 \lambda y_t^2 +4 \lambda \mbox{Tr}\Big({Y_\nu  Y_{\nu}^{\dagger}}\Big) \nonumber \\
 &-6 y_t^4  
 -2 \mbox{Tr}\Big({Y_\nu  Y_{\nu}^{\dagger}  Y_\nu  Y_{\nu}^{\dagger}}\Big) 
 \end{align}
 \begin{align}
& \beta_{\lambda}^{(2)}  =  
-\frac{3411}{2000} g_{1}^{6} -\frac{1677}{400} g_{1}^{4} g_{2}^{2} -\frac{289}{80} g_{1}^{2} g_{2}^{4} +\frac{305}{16} g_{2}^{6} +\frac{1887}{200} g_{1}^{4} \lambda +\frac{117}{20} g_{1}^{2} g_{2}^{2} \lambda -\frac{73}{8} g_{2}^{4} \lambda \nonumber \\
&+\frac{108}{5} g_{1}^{2} \lambda^{2} +108 g_{2}^{2} \lambda^{2} 
 -312 \lambda^{3}  
 -\frac{171}{100} g_{1}^{4} y_t^2 
 +\frac{63}{10} g_{1}^{2} g_{2}^{2} y_t^2 -\frac{9}{4} g_{2}^{4} y_t^2 
+\frac{17}{2} g_{1}^{2} \lambda y_t^2 +\frac{45}{2} g_{2}^{2} \lambda y_t^2 \nonumber \\ 
 &+80 g_{3}^{2} \lambda y_t^2 -144 \lambda^{2} y_t^2 -\frac{9}{100} g_{1}^{4} \mbox{Tr}\Big({Y_\nu  Y_{\nu}^{\dagger}}\Big) -\frac{3}{10} g_{1}^{2} g_{2}^{2} \mbox{Tr}\Big({Y_\nu  Y_{\nu}^{\dagger}}\Big)  
 -\frac{3}{4} g_{2}^{4} \mbox{Tr}\Big({Y_\nu  Y_{\nu}^{\dagger}}\Big)\\ \nonumber
& +\frac{3}{2} g_{1}^{2} \lambda \mbox{Tr}\Big({Y_\nu  Y_{\nu}^{\dagger}}\Big) +\frac{15}{2} g_{2}^{2} \lambda \mbox{Tr}\Big({Y_\nu  Y_{\nu}^{\dagger}}\Big) -48 \lambda^{2} \mbox{Tr}\Big({Y_\nu  Y_{\nu}^{\dagger}}\Big) 
   -\frac{8}{5} g_{1}^{2} y_t^4 
 -32 g_{3}^{2} y_t^4 -3 \lambda y_t^4 \\ \nonumber
&- \lambda \mbox{Tr}\Big({Y_\nu  Y_{\nu}^{\dagger}  Y_\nu  Y_{\nu}^{\dagger}}\Big)  
 +30 y_t^6 +10 \mbox{Tr}\Big({Y_\nu  Y_{\nu}^{\dagger}  Y_\nu  Y_{\nu}^{\dagger}  Y_\nu  Y_{\nu}^{\dagger}}\Big) 
\end{align}
\subsection{Yukawa Couplings}

The one-loop and two-loop RG corrections for the most relevant Yukawa couplings in the simplest inverse seesaw model are given by
{\allowdisplaybreaks  \begin{align} 
\beta_{Y_\nu}^{(1)} & =  
\frac{3}{2} {Y_\nu  Y_{\nu}^{\dagger}  Y_\nu} + Y_\nu \Big(3 y_t^2  -\frac{9}{20} g_{1}^{2}  -\frac{9}{4} g_{2}^{2} + \mbox{Tr}\Big({Y_\nu  Y_{\nu}^{\dagger}}\Big)\Big)\\ 
\beta_{Y_\nu}^{(2)} & =  
\frac{1}{80} \Big(279 g_{1}^{2} {Y_\nu  Y_{\nu}^{\dagger}  Y_\nu} +675 g_{2}^{2} {Y_\nu  Y_{\nu}^{\dagger}  Y_\nu} -960 \lambda {Y_\nu  Y_{\nu}^{\dagger}  Y_\nu} 
  +120 {Y_\nu  Y_{\nu}^{\dagger}  Y_\nu  Y_{\nu}^{\dagger}  Y_\nu} 
  -540 {Y_\nu  Y_{\nu}^{\dagger}  Y_\nu} y_t^2  \\ \nonumber
&-180 {Y_\nu  Y_{\nu}^{\dagger}  Y_\nu} \mbox{Tr}\Big({Y_\nu  Y_{\nu}^{\dagger}}\Big)  
 +2 Y_\nu \Big(21 g_{1}^{4} -54 g_{1}^{2} g_{2}^{2} -230 g_{2}^{4} +240 \lambda^{2}  
 +85 g_{1}^{2} y_t^2 +225 g_{2}^{2} y_t^2 \\ \nonumber
&+800 g_{3}^{2} y_t^2 +15 g_{1}^{2} \mbox{Tr}\Big({Y_\nu  Y_{\nu}^{\dagger}}\Big)  
 +75 g_{2}^{2} \mbox{Tr}\Big({Y_\nu  Y_{\nu}^{\dagger}}\Big)  
  -270 y_t^4 -90 \mbox{Tr}\Big({Y_\nu  Y_{\nu}^{\dagger}  Y_\nu  Y_{\nu}^{\dagger}}\Big) \Big)\Big)\\ 
\beta_{y_t}^{(1)} & =  
\frac{3}{2} y_t^3
 +y_t \Big(3 y_t^2  -8 g_{3}^{2}  -\frac{17}{20} g_{1}^{2}  -\frac{9}{4} g_{2}^{2}  + \mbox{Tr}\Big({Y_\nu  Y_{\nu}^{\dagger}}\Big)\Big)\\ 
\beta_{y_t}^{(2)} & =  
+\frac{1}{80} \Big(120 y_t^5 + y_t^3\Big(1280 g_{3}^{2} -180 \mbox{Tr}\Big({Y_\nu  Y_{\nu}^{\dagger}}\Big)  + 223 g_{1}^{2}  -540 y_t^2  + 675 g_{2}^{2}  -960 \lambda \Big)\nonumber \\ 
 &+y_t \Big(\frac{1187}{600} g_{1}^{4} -\frac{9}{20} g_{1}^{2} g_{2}^{2} -\frac{23}{4} g_{2}^{4} +\frac{19}{15} g_{1}^{2} g_{3}^{2} +9 g_{2}^{2} g_{3}^{2} -108 g_{3}^{4} +6 \lambda^{2}  
 +\frac{17}{8} g_{1}^{2} y_t^2 +\frac{45}{8} g_{2}^{2} y_t^2 \nonumber \\ 
 &+20 g_{3}^{2} y_t^2 +\frac{3}{8} g_{1}^{2} \mbox{Tr}\Big({Y_\nu  Y_{\nu}^{\dagger}}\Big) +\frac{15}{8} g_{2}^{2} \mbox{Tr}\Big({Y_\nu  Y_{\nu}^{\dagger}}\Big) 
  -\frac{27}{4} y_t^4 
 -\frac{9}{4} \mbox{Tr}\Big({Y_\nu  Y_{\nu}^{\dagger}  Y_\nu  Y_{\nu}^{\dagger}}\Big) \Big)
\end{align}} 
\section{RGEs: Inverse seesaw with majoron}
\label{app:inverse seesaw with majoron}

In the presence of the majoron the one- and two-loop RG corrections for the quartic scalar couplings in the inverse seesaw model are modified to
%
\subsection{Quartic scalar couplings}
{\allowdisplaybreaks  \begin{align} 
\label{lambdaphi}
\beta_{\lambda_\Phi}^{(1)} & =  
+\frac{27}{200} g_{1}^{4} +\frac{9}{20} g_{1}^{2} g_{2}^{2} +\frac{9}{8} g_{2}^{4} +\lambda_{\Phi\sigma}^{2}-\frac{9}{5} g_{1}^{2} \lambda_\Phi -9 g_{2}^{2} \lambda_\Phi +24 \lambda_\Phi^{2} 
 +12 \lambda_\Phi y_t^2 +4 \lambda_\Phi \mbox{Tr}\Big({Y_\nu  Y_{\nu}^{\dagger}}\Big)-6 y_t^4 \nonumber \\ 
 &-2 \mbox{Tr}\Big({Y_\nu  Y_{\nu}^{\dagger}  Y_\nu  Y_{\nu}^{\dagger}}\Big) \\ 
\beta_{\lambda_\Phi}^{(2)} & =  
-\frac{3411}{2000} g_{1}^{6} -\frac{1677}{400} g_{1}^{4} g_{2}^{2} -\frac{289}{80} g_{1}^{2} g_{2}^{4} +\frac{305}{16} g_{2}^{6} -4 \lambda_{\Phi\sigma}^{3} +\frac{1887}{200} g_{1}^{4} \lambda_\Phi +\frac{117}{20} g_{1}^{2} g_{2}^{2} \lambda_\Phi -\frac{73}{8} g_{2}^{4} \lambda_\Phi \\ \nonumber
&-10 \lambda_{\Phi\sigma}^{2} \lambda_\Phi 
 +\frac{108}{5} g_{1}^{2} \lambda_\Phi^{2} +108 g_{2}^{2} \lambda_\Phi^{2} -312 \lambda_\Phi^{3} 
  -4 \lambda_{\Phi\sigma}^{2} \mbox{Tr}\Big({Y_S  Y_S^*}\Big) 
 -\frac{171}{100} g_{1}^{4} y_t^2 +\frac{63}{10} g_{1}^{2} g_{2}^{2} y_t^2 \\ \nonumber
&-\frac{9}{4} g_{2}^{4} y_t^2 +\frac{17}{2} g_{1}^{2} \lambda_\Phi y_t^2 
 +\frac{45}{2} g_{2}^{2} \lambda_\Phi y_t^2 +80 g_{3}^{2} \lambda_\Phi y_t^2 -144 \lambda_\Phi^{2} y_t^2 -\frac{9}{100} g_{1}^{4} \mbox{Tr}\Big({Y_\nu  Y_{\nu}^{\dagger}}\Big) \nonumber \\ 
 &-\frac{3}{10} g_{1}^{2} g_{2}^{2} \mbox{Tr}\Big({Y_\nu  Y_{\nu}^{\dagger}}\Big) -\frac{3}{4} g_{2}^{4} \mbox{Tr}\Big({Y_\nu  Y_{\nu}^{\dagger}}\Big) +\frac{3}{2} g_{1}^{2} \lambda_\Phi \mbox{Tr}\Big({Y_\nu  Y_{\nu}^{\dagger}}\Big) +\frac{15}{2} g_{2}^{2} \lambda_\Phi \mbox{Tr}\Big({Y_\nu  Y_{\nu}^{\dagger}}\Big) 
 -48 \lambda_\Phi^{2} \mbox{Tr}\Big({Y_\nu  Y_{\nu}^{\dagger}}\Big) \nonumber \\ 
 &-\frac{8}{5} g_{1}^{2} y_t^4 -32 g_{3}^{2} y_t^4 -3 \lambda_\Phi y_t^4 - \lambda_\Phi \mbox{Tr}\Big({Y_\nu  Y_{\nu}^{\dagger}  Y_\nu  Y_{\nu}^{\dagger}}\Big) 
  +30 y_t^6 +10 \mbox{Tr}\Big({Y_\nu  Y_{\nu}^{\dagger}  Y_\nu  Y_{\nu}^{\dagger}  Y_\nu  Y_{\nu}^{\dagger}}\Big) \nonumber \\
\beta_{\lambda_{\Phi\sigma}}^{(1)} & =  
\frac{1}{10} \lambda_{\Phi\sigma} \Big(-9 g_{1}^{2} -45 g_{2}^{2} +40 \lambda_{\Phi\sigma} +80 \lambda_{\sigma} +120 \lambda_\Phi   +40 \mbox{Tr}\Big({Y_S  Y_S^*}\Big) +60 y_t^2  
 +20 \mbox{Tr}\Big({Y_\nu  Y_{\nu}^{\dagger}}\Big) \Big) \\
\beta_{\lambda_{\Phi\sigma}}^{(2)} & =  
+\frac{1671}{400} g_{1}^{4} \lambda_{\Phi\sigma} +\frac{9}{8} g_{1}^{2} g_{2}^{2} \lambda_{\Phi\sigma} -\frac{145}{16} g_{2}^{4} \lambda_{\Phi\sigma} +\frac{3}{5} g_{1}^{2} \lambda_{\Phi\sigma}^{2} +3 g_{2}^{2} \lambda_{\Phi\sigma}^{2} -11 \lambda_{\Phi\sigma}^{3} -48 \lambda_{\Phi\sigma}^{2} \lambda_{\sigma} \nonumber \\
&-40 \lambda_{\Phi\sigma} \lambda_{\sigma}^{2} 
 +\frac{72}{5} g_{1}^{2} \lambda_{\Phi\sigma} \lambda_\Phi +72 g_{2}^{2} \lambda_{\Phi\sigma} \lambda_\Phi -72 \lambda_{\Phi\sigma}^{2} \lambda -60 \lambda_{\Phi\sigma} \lambda_\Phi^{2} -8 \lambda_{\Phi\sigma}^{2} \mbox{Tr}\Big({Y_S  Y_S^*}\Big) \nonumber \\
& -32 \lambda_{\Phi\sigma} \lambda_{\sigma} \mbox{Tr}\Big({Y_S  Y_S^*}\Big) 
 +\frac{17}{4} g_{1}^{2} \lambda_{\Phi\sigma} y_t^2 +\frac{45}{4} g_{2}^{2} \lambda_{\Phi\sigma} y_t^2 +40 g_{3}^{2} \lambda_{\Phi\sigma} y_t^2 -12 \lambda_{\Phi\sigma}^{2} y_t^2 \nonumber \\ 
 &-72 \lambda_{\Phi\sigma} \lambda_\Phi y_t^2 +\frac{3}{4} g_{1}^{2} \lambda_{\Phi\sigma} \mbox{Tr}\Big({Y_\nu  Y_{\nu}^{\dagger}}\Big) +\frac{15}{4} g_{2}^{2} \lambda_{\Phi\sigma} \mbox{Tr}\Big({Y_\nu  Y_{\nu}^{\dagger}}\Big) -4 \lambda_{\Phi\sigma}^{2} \mbox{Tr}\Big({Y_\nu  Y_{\nu}^{\dagger}}\Big) \nonumber \\ 
 &-24 \lambda_{\Phi\sigma} \lambda_\Phi \mbox{Tr}\Big({Y_\nu  Y_{\nu}^{\dagger}}\Big)  
  -24 \lambda_{\Phi\sigma} \mbox{Tr}\Big({Y_S  Y_S^*  Y_S  Y_S^*}\Big) -\frac{27}{2} \lambda_{\Phi\sigma} y_t^4 -\frac{9}{2} \lambda_{\Phi\sigma} \mbox{Tr}\Big({Y_\nu  Y_{\nu}^{\dagger}  Y_\nu  Y_{\nu}^{\dagger}}\Big) \\ 
\beta_{\lambda_{\sigma}}^{(1)} & =  
2 \Big(10 \lambda_{\sigma}^{2}  + 4 \lambda_{\sigma} \mbox{Tr}\Big({Y_S  Y_S^*}\Big)  -8 \mbox{Tr}\Big({Y_S  Y_S^*  Y_S  Y_S^*}\Big)  + \lambda_{\Phi\sigma}^{2}\Big)\\ 
\beta_{\lambda_{\sigma}}^{(2)} & =  
+\frac{12}{5} g_{1}^{2} \lambda_{\Phi\sigma}^{2} +12 g_{2}^{2} \lambda_{\Phi\sigma}^{2} -8 \lambda_{\Phi\sigma}^{3} -20 \lambda_{\Phi\sigma}^{2} \lambda_{\sigma} -240 \lambda_{\sigma}^{3} 
 -80 \lambda_{\sigma}^{2} \mbox{Tr}\Big({Y_S  Y_S^*}\Big) -12 \lambda_{\Phi\sigma}^{2} y_t^2 \nonumber \\
& -4 \lambda_{\Phi\sigma}^{2} \mbox{Tr}\Big({Y_\nu  Y_{\nu}^{\dagger}}\Big) +16 \lambda_{\sigma} \mbox{Tr}\Big({Y_S  Y_S^*  Y_S  Y_S^*}\Big) +256 \mbox{Tr}\Big({Y_S  Y_S^*  Y_S  Y_S^*  Y_S  Y_S^*}\Big) 
\end{align}}

\vfill
\subsection{Yukawa Couplings}
Likewise, in the presence of the majoron the one- and two-loop RG corrections for the Yukawas in the inverse seesaw model are modified to
{\allowdisplaybreaks  \begin{align} 
\beta_{Y_\nu}^{(1)} & =  
\frac{3}{2} {Y_\nu  Y_{\nu}^{\dagger}  Y_\nu}  
 +Y_\nu \Big(3 y_t^2  -\frac{9}{20} g_{1}^{2}  -\frac{9}{4} g_{2}^{2}  + \mbox{Tr}\Big({Y_\nu  Y_{\nu}^{\dagger}}\Big)\Big)\\ 
\beta_{Y_\nu}^{(2)} & =  
\frac{1}{80} \Big(279 g_{1}^{2} {Y_\nu  Y_{\nu}^{\dagger}  Y_\nu} +675 g_{2}^{2} {Y_\nu  Y_{\nu}^{\dagger}  Y_\nu} -960 \lambda_\Phi {Y_\nu  Y_{\nu}^{\dagger}  Y_\nu}  
  +120 {Y_\nu  Y_{\nu}^{\dagger}  Y_\nu  Y_{\nu}^{\dagger}  Y_\nu} -540 {Y_\nu  Y_{\nu}^{\dagger}  Y_\nu} y_t^2 \nonumber \\
&-180 {Y_\nu  Y_{\nu}^{\dagger}  Y_\nu} \mbox{Tr}\Big({Y_\nu  Y_{\nu}^{\dagger}}\Big)  
 +2 Y_\nu \Big(21 g_{1}^{4} -54 g_{1}^{2} g_{2}^{2} -230 g_{2}^{4} +20 \lambda_{\Phi\sigma}^{2} +240 \lambda_\Phi^{2}  
  +85 g_{1}^{2} y_t^2 \nonumber \\
&+225 g_{2}^{2} y_t^2 +800 g_{3}^{2} y_t^2 
 +15 g_{1}^{2} \mbox{Tr}\Big({Y_\nu  Y_{\nu}^{\dagger}}\Big) +75 g_{2}^{2} \mbox{Tr}\Big({Y_\nu  Y_{\nu}^{\dagger}}\Big)    
   -270 y_t^4 -90 \mbox{Tr}\Big({Y_\nu  Y_{\nu}^{\dagger}  Y_\nu  Y_{\nu}^{\dagger}}\Big) \Big)\Big)\\ 
\beta_{y_t}^{(1)} & =  
\frac{3}{2} y_t^3 + y_t \Big( 3 y_t^2  -8 g_{3}^{2}  -\frac{17}{20} g_{1}^{2}  -\frac{9}{4} g_{2}^{2}   + \mbox{Tr}\Big({Y_\nu  Y_{\nu}^{\dagger}}\Big)\Big)\\ 
\beta_{y_t}^{(2)} & =  
+\frac{1}{80} \Big(120 y_t^5 + y_t^3 \Big(1280 g_{3}^{2}    -180 \mbox{Tr}\Big({Y_\nu  Y_{\nu}^{\dagger}}\Big)  + 223 g_{1}^{2}    -540 y_t^2  + 675 g_{2}^{2}  -960 \lambda_\Phi \Big)\nonumber \\ 
 &+y_t \Big(\frac{1187}{600} g_{1}^{4} -\frac{9}{20} g_{1}^{2} g_{2}^{2} -\frac{23}{4} g_{2}^{4} +\frac{19}{15} g_{1}^{2} g_{3}^{2} +9 g_{2}^{2} g_{3}^{2} -108 g_{3}^{4} +\frac{1}{2} \lambda_{\Phi\sigma}^{2} +6 \lambda_\Phi^{2}  +\frac{17}{8} g_{1}^{2} y_t^2 \\ \nonumber
& +\frac{45}{8} g_{2}^{2} y_t^2
 +20 g_{3}^{2} y_t^2 +\frac{3}{8} g_{1}^{2} \mbox{Tr}\Big({Y_\nu  Y_{\nu}^{\dagger}}\Big) +\frac{15}{8} g_{2}^{2} \mbox{Tr}\Big({Y_\nu  Y_{\nu}^{\dagger}}\Big)  
    -\frac{27}{4} y_t^4 -\frac{9}{4} \mbox{Tr}\Big({Y_\nu  Y_{\nu}^{\dagger}  Y_\nu  Y_{\nu}^{\dagger}}\Big) \Big)\\ 
\beta_{Y_S}^{(1)} & =  
2 Y_S \mbox{Tr}\Big({Y_S  Y_S^*}\Big)  + 4 {Y_S  Y_S^*  Y_S} \\ 
\beta_{Y_S}^{(2)} & =  
28 {Y_S  Y_S^*  Y_S  Y_S^*  Y_S}  -4 {Y_S  Y_S^*  Y_S} \Big(3 \mbox{Tr}\Big({Y_S  Y_S^*}\Big)  + 8 \lambda_{\sigma} \Big) + Y_S \Big(-12 \mbox{Tr}\Big({Y_S  Y_S^*  Y_S  Y_S^*}\Big) \nonumber \\
& + 4 \lambda_{\sigma}^{2}  + \lambda_{\Phi\sigma}^{2}\Big)
\end{align}} 
\acknowledgments
  This work is supported by the Spanish grant FPA2017-85216-P (AEI/FEDER, UE), PROMETEO/2018/165 (Generalitat Valenciana),
  Funda\c{c}{\~a}o para a Ci{\^e}ncia e a Tecnologia (FCT, Portugal) under project CERN/FIS-PAR/0004/2019, and the Spanish Red Consolider MultiDark FPA2017-90566-REDC.
\bibliographystyle{JHEP}
\bibliography{bibliography} 
\end{document}